%Paper: hep-th/9302055
%From: grosche@tsmi19.sissa.it
%Date: Sat, 13 Feb 1993 14:03:26 +0100

%----------------------------------------------------------------------
%                       Use AmSTeX
%----------------------------------------------------------------------
\input amstex
%%%%%%%%%%%%%%%%%%%%%%%%%%%%%%%%%%%%%%%%%%%%%%%%%%%%%%%%%%%%%%%%%%%%%%%
%
% First we have a character check
%
% ! exclamation mark    " double quote
% # hash                ` opening quote (grave)
% & ampersand           ' closing quote (acute)
% $ dollar              % percent
% ( open parenthesis    ) close paren.
% - hyphen              = equals sign
% | vertical bar        ~ tilde
% @ at sign             _ underscore
% { open curly brace    } close curly
% [ open square         ] close square bracket
% + plus sign           ; semi-colon
% * asterisk            : colon
% < open angle bracket  > close angle
% , comma               . full stop
% ? question mark       / forward slash
% \ backslash           ^ circumflex
%
% ABCDEFGHIJKLMNOPQRSTUVWXYZ
% abcdefghijklmnopqrstuvwxyz
% 1234567890
%
%%%%%%%%%%%%%%%%%%%%%%%%%%%%%%%%%%%%%%%%%%%%%%%%%%%%%%%%%%%%%%%%%%%%%

 \magnification=1200
\hsize=31pc
\vsize=55 truepc
\hfuzz=2pt
\vfuzz=4pt
%\pretolerance=5000
%\tolerance=5000
\pretolerance=500
\tolerance=500
\parskip=0pt plus 1pt
\parindent=16pt
%

%%%%%%%%%%%%%%%%%%%%%%%%% Font definitions %%%%%%%%%%%%%%%%%%%%%%%%%%
%
% Fonts for title
%
\font\fourteenrm=cmr10 scaled \magstep2
\font\fourteeni=cmmi10 scaled \magstep2
\font\fourteenbf=cmbx10 scaled \magstep2
\font\fourteenit=cmti10 scaled \magstep2
\font\fourteensy=cmsy10 scaled \magstep2

% Font for small caps within title and authors names
%
\font\large=cmbx10 scaled \magstep1

% Font for matrices (please replace with cmbx10
% if you do not have this font available)
%
\font\sans=cmssbx10

% Font for matrices (use \bss{x} for bold sans serif within maths)
%

% Font for vectors (bold italic). Please replace with cmbx10
% if you do not have this font available.
% Use \bi{r} for bold italic r within maths
%

% Fonts for small type (if used)
%
\font\eightrm=cmr8
\font\eighti=cmmi8
\font\eightbf=cmbx8
\font\eightit=cmti8

\font\eightsy=cmsy8
\font\sixrm=cmr6
\font\sixi=cmmi6
\font\sixsy=cmsy6

%%%% Definitions of tenpoint, eightpoint and fourteenpoint families
%
\def\tenpoint{\def\rm{\fam0\tenrm}%
  \textfont0=\tenrm \scriptfont0=\sevenrm
                      \scriptscriptfont0=\fiverm
  \textfont1=\teni  \scriptfont1=\seveni
                      \scriptscriptfont1=\fivei
  \textfont2=\tensy \scriptfont2=\sevensy
                      \scriptscriptfont2=\fivesy
  \textfont3=\tenex   \scriptfont3=\tenex
                      \scriptscriptfont3=\tenex
  \textfont\itfam=\tenit  \def\it{\fam\itfam\tenit}%
  \textfont\slfam=\tensl  \def\sl{\fam\slfam\tensl}%
  \textfont\bffam=\tenbf  \scriptfont\bffam=\sevenbf
                            \scriptscriptfont\bffam=\fivebf
                            \def\bf{\fam\bffam\tenbf}%
  \normalbaselineskip=20 truept
  \setbox\strutbox=\hbox{\vrule height14pt depth6pt
width0pt}%
  \let\sc=\eightrm \normalbaselines\rm}
\def\eightpoint{\def\rm{\fam0\eightrm}%
  \textfont0=\eightrm \scriptfont0=\sixrm
                      \scriptscriptfont0=\fiverm
  \textfont1=\eighti  \scriptfont1=\sixi
                      \scriptscriptfont1=\fivei
  \textfont2=\eightsy \scriptfont2=\sixsy
                      \scriptscriptfont2=\fivesy
  \textfont3=\tenex   \scriptfont3=\tenex
                      \scriptscriptfont3=\tenex
  \textfont\itfam=\eightit  \def\it{\fam\itfam\eightit}%
  \textfont\bffam=\eightbf  \def\bf{\fam\bffam\eightbf}%
  \normalbaselineskip=16 truept
  \setbox\strutbox=\hbox{\vrule height11pt depth5pt width0pt}}
\def\fourteenpoint{\def\rm{\fam0\fourteenrm}%
  \textfont0=\fourteenrm \scriptfont0=\tenrm
                      \scriptscriptfont0=\eightrm
  \textfont1=\fourteeni  \scriptfont1=\teni
                      \scriptscriptfont1=\eighti
  \textfont2=\fourteensy \scriptfont2=\tensy
                      \scriptscriptfont2=\eightsy
  \textfont3=\tenex   \scriptfont3=\tenex
                      \scriptscriptfont3=\tenex
  \textfont\itfam=\fourteenit  \def\it{\fam\itfam\fourteenit}%
  \textfont\bffam=\fourteenbf  \scriptfont\bffam=\tenbf
                             \scriptscriptfont\bffam=\eightbf
                             \def\bf{\fam\bffam\fourteenbf}%
  \normalbaselineskip=24 truept
  \setbox\strutbox=\hbox{\vrule height17pt depth7pt width0pt}%
  \let\sc=\tenrm \normalbaselines\rm}
\def\today{\number\day\ \ifcase\month\or
  January\or February\or March\or April\or May\or June\or
  July\or August\or September\or October\or November\or
December\fi
  \space \number\year}
\def\monthyear{\ifcase\month\or
  January\or February\or March\or April\or May\or June\or
  July\or August\or September\or October\or November\or
December\fi
  \space \number\year}

%%%%%%%%%%%%%%%%%%%%%%%% Counter definitions %%%%%%%%%%%%%%%%%%%%%%%%
%
\newcount\secno      %section number
\newcount\subno      %number of subsection
\newcount\subsubno   %number of subsubsection
\newcount\appno      %appendix number
\newcount\tableno    %table number
\newcount\figureno   %figure number
%

%%%%%%%%%%%%%%%%%%%%%%%%%%%%% Baselineskip %%%%%%%%%%%%%%%%%%%%%%%%%%
%
\normalbaselineskip=20 truept
\baselineskip=20 truept

%%%%%%%%%%%%%%%%%%%% Specific formatting commands %%%%%%%%%%%%%%%%%%%
%
% Title of article
%
\def\title#1
   {\vglue1truein
   {\baselineskip=24 truept
    \pretolerance=10000
    \raggedright
    \noindent \fourteenpoint\bf #1\par}
    \vskip1truein minus36pt}
%

% Author names
% (The names of all the authors should be in the form initials then
% surname. There should be no points after initials.)
%
\def\author#1
  {{\pretolerance=10000
    \raggedright
    \noindent {\large #1}\par}}

% Address of the authors
% (If authors are at differing addresses use one \address for each)
%
\def\address#1
   {\bigskip
    \noindent \rm #1\par}

% Short title (not more than fifty characters)
%
\def\shorttitle#1
   {\vfill
    \noindent \rm Short title: {\sl #1}\par
    \medskip}

% Physics Abstracts classification numbers
%
\def\pacs#1
   {\noindent \rm PACS number(s): #1\par
    \medskip}

% Journal article submitted to
%
\def\jnl#1
   {\noindent \rm Submitted to: {\sl #1}\par
    \medskip}

% Today's date
%
\def\date
   {\noindent Date: \today\par
    \medskip}

% Start of abstract
%

% Keyword abstract - only required for J. Phys. G
%
\def\keyword#1
   {\bigskip
    \noindent {\bf Keyword abstract: }\rm#1}

% End of abstract
%

% Contents page (only required for Reports on Progress in Physics)
%
% Heading for contents page
%

% Entry in list of contents (section headings)
%
\def\entry#1#2#3
   {\noindent
    \hangindent=20pt
    \hangafter=1
    \hbox to20pt{#1 \hss}#2\hfill #3\par}

% Subentry in list of contents (subsection heading).
% (Subsubsection headings do not appear in the contents list)
%
\def\subentry#1#2#3
   {\noindent
    \hangindent=40pt
    \hangafter=1
    \hskip20pt\hbox to20pt{#1 \hss}#2\hfill #3\par}
\def\checkforsub{\futurelet\nexttok\decide}
\def\ssf{\relax}
\def\decide{\if\nexttok\ssf\let\endspace=\nospace
                \else\let\endspace=\extraspace\fi\endspace}
\def\nospace{\nobreak\par\nobreak}
%
% Section heading (#1 is title of section, no number required)
%
\def\section#1{%
    \goodbreak
    \vskip24pt plus12pt minus12pt
    \nobreak
    \gdef\extraspace{\nobreak\bigskip\noindent\ignorespaces}%
    \noindent
    \subno=0 \subsubno=0
    \global\advance\secno by 1
    \noindent {\bf \the\secno. #1}\par\checkforsub}

% Subsection heading (#1 is title of subsection, no number required)
%
\def\subsection#1{%
     \goodbreak
     \vskip24pt plus12pt minus6pt
     \nobreak
     \gdef\extraspace{\nobreak\medskip\noindent\ignorespaces}%
     \noindent
     \subsubno=0
     \global\advance\subno by 1
     \noindent {\sl \the\secno.\the\subno. #1\par}\checkforsub}

% Subsubsection heading (#1 is title of subsubsection,
% no number is required)
%
\def\subsubsection#1{%
     \goodbreak
     \vskip15pt plus6pt minus6pt
     \nobreak\noindent
     \global\advance\subsubno by 1
     \noindent {\sl \the\secno.\the\subno.\the\subsubno. #1}\null.
     \ignorespaces}

% Heading for an appendix, #1 is title of appendix,
% no number or letter required
%
\def\appendix#1
   {\vskip0pt plus.1\vsize\penalty-250
    \vskip0pt plus-.1\vsize\vskip24pt plus12pt minus6pt
    \subno=0
    \global\advance\appno by 1
    \noindent {\bf Appendix \the\appno. #1\par}
    \bigskip
    \noindent}

% Heading for subsection within an appendix
%
\def\subappendix#1
   {\vskip-\lastskip
    \vskip36pt plus12pt minus12pt
    \bigbreak
    \global\advance\subno by 1
    \noindent {\sl \the\appno.\the\subno. #1\par}
    \nobreak
    \medskip
    \noindent}

% Heading for acknowledgments
%

%%%%%%%%%%%%%%%%%%%%%%%%%Macros for Tables%%%%%%%%%%%%%%%%%%%%%%%%%%%%

% Heading for start of tables section
%

% Table caption. #1 is caption, no number required
%
\def\tabcaption#1
   {\global\advance\tableno by 1
    \noindent {\bf Table \the\tableno.} \rm#1\par
    \bigskip}

% Definition of boldrule and medrule
%

% The halign (actually ialign) command for tables
% (THIS MUST BE COPIED FOR EACH TABLE---WITHOUT THE PER CENT SIGN!)
%
% \ialign{#\hfil&&\hglue 2pc plus2pc minus1pc#\hfil\cr

% A small negative skip for use in tables
%

% A macro for a footnote to a table
%

% Heading for list of figure captions
%

% Figure caption, #1 is caption, no number required
%
\def\figcaption#1
   {\global\advance\figureno by 1
    \noindent {\bf Figure \the\figureno.} \rm#1\par
    \bigskip}

% Heading for list of references
%

% Heading for list of numbered references
%

% Reference to a journal article in Harvard (alphabetical) system
%
\def\refjl#1#2#3#4
   {\hangindent=16pt
    \hangafter=1
    \rm #1
   {\frenchspacing\sl #2
    \bf #3}
    #4\par}

% Reference to a book or report in Harvard (alphabetical) system
%
\def\refbk#1#2#3
   {\hangindent=16pt
    \hangafter=1
    \rm #1
   {\frenchspacing\sl #2}
    #3\par}

% Reference to a journal article in numerical system
%
\def\numrefjl#1#2#3#4#5
   {\parindent=40pt
    \hang
    \noindent
    \rm {\hbox to 30truept{\hss #1\quad}}#2
   {\frenchspacing\sl #3\/
    \bf #4}
    #5\par\parindent=16pt}

% Reference to a book or report in numerical system
%
\def\numrefbk#1#2#3#4
   {\parindent=40pt
    \hang
    \noindent
    \rm {\hbox to 30truept{\hss #1\quad}}#2
   {\frenchspacing\sl #3\/}
    #4\par\parindent=16pt}

% Dash for use with repeated authors in reference lists
%

\def\ref#1{\par\noindent \hbox to 21pt{\hss
#1\quad}\frenchspacing\ignorespaces}

% Fraction, alternative to \over
%
\def\frac#1#2{{#1 \over #2}}

% Renaming the dot under macro
%

% \d now used for differential d in mathematics
%

% \e gives roman e for exponential e in mathematics
%
\def\e{\operatorname{e}}
%\def\e{\hbox{\rm e}}

% \i gives roman i for square root of minus one in maths mode
% and \ii used for dotless i in text mode

%\def\i{\hbox{\rm i}}
\def\i{\operatorname{i}}
\chardef\ii="10

% Small (text size) fraction within displayed mathematics
%

% et al
%

% Redefinition of footnote macros to lose rule and remove indentation
%

\catcode`\@=11
\def\vfootnote#1{\insert\footins\bgroup
    \interlinepenalty=\interfootnotelinepenalty
    \splittopskip=\ht\strutbox % top baseline for broken footnotes
    \splitmaxdepth=\dp\strutbox \floatingpenalty=20000
    \leftskip=0pt \rightskip=0pt \spaceskip=0pt \xspaceskip=0pt
    \noindent\eightpoint\rm #1\ \ignorespaces\footstrut\futurelet\next\fo@t}

% Special macros for display equations
%
% \eq(#1) will give the equation number (#1) on the right
% instead of \eqno
%
\def\eq(#1){\hfill\llap{(#1)}}
\catcode`\@=12
%
% Macro for special accented characters
%
% Vectors with hats

% vectors with overbar

% roman characters with right pointing arrow

%
% Abbreviations for IOPP journals
%

        %1968-87
   %1988 and onwards
     %1968--1988
        %1989 and onwards

         %1968-89

           %1975--1988
     %1989 and onwards

         %1989 and onwards

%
% Other commonly quoted journals
%

%
% Miscellaneous definitions
%
% Bold nabla

%
% Bold dot for vector dot products
%

%
% Small space between lines in alignments or displayed maths

%
% Half line space for within tables or alignments

%
% Small negative space to close up lines above rules in tables

%
% greater than approximately signs
% \def\gap{\;\amex{\char'046}\;}           % use these if ams
%  extension fonts available
% \def\sgap{\;\hbox{\samsx \char'046}\;}   % see above
\def\gap{\;\lower3pt\hbox{$\buildrel > \over \sim$}\;}
%
% less than approximately
%
% \def\lap{\;\amex{\char'056}\;}
% \def\slap{\;\hbox{\samsx \char'056}\;}
\def\lap{\;\lower3pt\hbox{$\buildrel < \over \sim$}\;}
% space between parts of short equations
\def\tqs{\hbox to 25pt{\hfil}}

   %order of
%\def\Cal#1{{\cal #1}}

{\obeylines\gdef\startdisplay#1
  {\catcode`\^^M=5$$#1\halign\bgroup\indent##\hfil&&\qquad##\hfil\cr}}
\outer\def\enddisplay{\crcr\egroup$$}

\chardef\other=12
\def\ttverbatim{\begingroup \catcode`\\=\other \catcode`\{=\other
  \catcode`\}=\other \catcode`\$=\other \catcode`\&=\other
  \catcode`\#=\other \catcode`\%=\other \catcode`\~=\other
  \catcode`\_=\other \catcode`\^=\other
  \obeyspaces \obeylines \tt}
{\obeyspaces\gdef {\ }}  % \obeyspaces now gives \ , not \space

\outer\def\begintt{$$\let\par=\endgraf \ttverbatim \parskip=0pt
  \catcode`\|=0 \rightskip=-5pc \ttfinish}
{\catcode`\|=0 |catcode`|\=\other % | is temporary escape character
  |obeylines % end of line is active
  |gdef|ttfinish#1^^M#2\endtt{#1|vbox{#2}|endgroup$$}}

\catcode`\|=\active
{\obeylines\gdef|{\ttverbatim\spaceskip=.5em plus.25em minus.15em
                                            \let^^M=\ \let|=\endgroup}}%

\TagsOnRight
%----------------------------------------------------------------------
%                       For Preprint
%----------------------------------------------------------------------

\tracingstats=1    % Speicherplatzstatistik

\font\twelverm=cmr10 scaled 1200

\normalbaselineskip=15pt
\baselineskip=14pt

%----------------------------------------------------------------------
%                        DEFINITIONS
%----------------------------------------------------------------------
\def\CD{{\Cal D}}
\def\CL{{\Cal L}}

\def\GV{G^{(V)}}
\def\GB{G^{(Box)}}
\def\GWa{G^{(Wall)}}

\def\KW{K^{(W)}}

\def\Gd{G^{(\delta)}}
\def\Kd{K^{(\delta)}}

\def\ih{{\i\over\hbar}}
\def\overh{{1\over\hbar}}
\def\bbbr{\operatorname{{I\!R}}}                     %reelle Zahlen
\def\bbbn{\operatorname{{I\!N}}}                     %natuerliche Zahlen
\def\Ai{\operatorname{Ai}}
\def\Bi{\operatorname{Bi}}
\font\sans=cmssbx10
\def\sf{\sans}
\def\bbbz{{\mathchoice {\hbox{$\sf\textstyle Z\kern-0.4em Z$}}
{\hbox{$\sf\textstyle Z\kern-0.4em Z$}}
{\hbox{$\sf\scriptstyle Z\kern-0.3em Z$}}
{\hbox{$\sf\scriptscriptstyle Z\kern-0.2em Z$}}}}    %ganze Zahlen
\def\AA{{\hbox{\rm A}\NUM}}
\def\BB{{\hbox{\rm A}\NUM}}

\def\viert{{1\over4}}
\def\half{{1\over2}}

\def\erfc{\operatorname{erfc}}

\def\myalign{\allowdisplaybreaks\align}
\def\dfrac{\dsize\frac}
\def\pha{\phantom{.}}
\hfuzz=3pt

\newcount\Chapno
\def\PLUS{\advance\Chapno by 1}
\def\NUM{\the\Chapno}
\Chapno=1

\newcount\glno
\def\plus{\advance\glno by 1}
\def\num{\the\glno}

\newcount\Refno
\def\add{\advance\Refno by 1}
\Refno=1

\edef\AGHH{\the\Refno}\add
\edef\ANIKZ{\the\Refno}\add
\edef\AGS{\the\Refno}\add
\edef\BAVE{\the\Refno}\add
\edef\BVK{\the\Refno}\add
\edef\BAU{\the\Refno}\add
\edef\BHR{\the\Refno}\add
\edef\CIWI{\the\Refno}\add
\edef\CAR{\the\Refno}\add
\edef\CFG{\the\Refno}\add
\edef\CCH{\the\Refno}\add
\edef\CHc{\the\Refno}\add
\edef\CMS{\the\Refno}\add
\edef\COFRI{\the\Refno}\add
\edef\DMN{\the\Refno}\add
\edef\DURa{\the\Refno}\add
\edef\DURd{\the\Refno}\add
\edef\DURe{\the\Refno}\add
\edef\DURf{\the\Refno}\add
\edef\EMOTa{\the\Refno}\add
\edef\EMOTb{\the\Refno}\add
\edef\FALK{\the\Refno}\add
\edef\FAKL{\the\Refno}\add
\edef\FARI{\the\Refno}\add
\edef\FH{\the\Refno}\add
\edef\FLU{\the\Refno}\add
\edef\GASCH{\the\Refno}\add
\edef\GOO{\the\Refno}\add
\edef\GOBR{\the\Refno}\add
\edef\GRA{\the\Refno}\add
\edef\GROb{\the\Refno}\add
\edef\GROc{\the\Refno}\add
\edef\GROe{\the\Refno}\add
\edef\GROh{\the\Refno}\add
\edef\GROm{\the\Refno}\add
\edef\GROu{\the\Refno}\add
\edef\GROw{\the\Refno}\add
\edef\GRSa{\the\Refno}\add
\edef\GRSb{\the\Refno}\add
\edef\GRSf{\the\Refno}\add
\edef\GRSg{\the\Refno}\add
\edef\INO{\the\Refno}\add
\edef\INOSa{\the\Refno}\add
\edef\INOSb{\the\Refno}\add
\edef\JAKL{\the\Refno}\add
\edef\KLRO{\the\Refno}\add
\edef\KLE{\the\Refno}\add
\edef\KLEMUS{\the\Refno}\add
\edef\KOM{\the\Refno}\add
\edef\LABH{\the\Refno}\add
\edef\MCG{\the\Refno}\add
\edef\MCGHU{\the\Refno}\add
\edef\MOS{\the\Refno}\add
\edef\MACR{\the\Refno}\add
\edef\PAKSa{\the\Refno}\add
\edef\PI{\the\Refno}\add
\edef\SCHU{\the\Refno}\add
\edef\SIEB{\the\Refno}\add
\edef\STEc{\the\Refno}\add
\edef\STEa{\the\Refno}\add
\edef\STEb{\the\Refno}\add
\edef\VAW{\the\Refno}\add
\edef\WIE{\the\Refno}\add
\edef\ZHCH{\the\Refno}\add

%----------------------------------------------------------------------
%                          END OF FILEM
%----------------------------------------------------------------------

{\nopagenumbers
\pageno=0
\centerline{February 1993\hfill SISSA/18/93/FM}
\vskip1cm
\centerline{\fourteenpoint $\delta$-FUNCTION PERTURBATIONS AND}
\bigskip
\centerline{\fourteenpoint BOUNDARY PROBLEMS BY PATH INTEGRATION}
\vskip1cm
\centerline{\twelverm CHRISTIAN GROSCHE}
\bigskip
\centerline{\it Scuola Internazionale Superiore di Studi Avanzati}
\centerline{\it International School for Advanced Studies}
\centerline{\it Via Beirut 4, 34014 Trieste, Miramare, Italy}
\vfill
\midinsert
\narrower
\noindent
{\bf Abstract.}
A wide class of boundary problems in quantum mechanics is discussed by
using path integrals. This includes motion in half-spaces, radial boxes,
rings, and moving boundaries. As a preparation the formalism for the
incorporation of $\delta$-function perturbations is outlined, which
includes the discussion of multiple $\delta$-function perturbations,
$\delta$-function perturbations along perpendicular lines and planes,
and moving $\delta$-function perturbations. The limiting process, where
the strength of the $\delta$-function perturbations gets infinite
repulsive, has the effect of producing impenetrable walls at the
locations of the $\delta$-function perturbations, i.e.\ a consistent
description for boundary problems with Dirichlet boundary-condition
emerges. Several examples illustrate the formalism.
\endinsert
\eject}
\pageno=1

%----------------------------------------------------------------------
%                          END OF FILE0
%----------------------------------------------------------------------

\glno=0                      %I
\section{Introduction}
Boundary problems, respectively boundary-value problems, appear at
almost every stage of quantum physics. In a conventional treatment
boundary-conditions are usually used to define the solution of a
particular solution, say of the Schr\"odinger equation, in an
unambiguous way. Such boundary-conditions, e.g. vanishing
wave-functions at infinity, are the basic requirement for the very
set-up of the relevant Hilbert space.

Boundary-conditions at infinity, however, are a very specific
idealization and for many more physical situation not appropriate.
Typical experimental situations require boundary-conditions at a
finite distance from the origin, for instance in electrodynamics
Dirichlet or Neumann boundary-conditions are the most discussed.
Analogous considerations in quantum mechanics lead to Dirichlet and
Neumann boundary-conditions, respectively, for the wave functions. In
the following we will consider such quantum mechanical problems
with Dirichlet boundary-conditions.

To deal with these problems, there have been only few attempts in the
literature. For instance, numerical studies can be found in
Refs.~[\ANIKZ, \COFRI, \MACR, \VAW] for the problem of an oscillator
in a box.

Studies of magnetic systems in euclidean and radial boxes in the
context a solid state physics are due to Falkovsky and Klama [\FALK,
\FAKL], and recently by Klama and R\"ossler [\KLRO]. In these studies
the authors succeeded in the explicit construction of the relevant
Green function by exploiting the given boundary-conditions for a
particular system. However, their approach lacks a systematic approach
to the whole problem of boundary-conditions, in particular for
multidimensional boundary problems with several boundaries. Also, path
integral considerations were not made.

Path integral studies for boundary problems have been done by several
authors. First, by the explicit construction of the propagator by using
the ``mirror'' principle (see below), c.f.\ in the classical textbooks
on path integrals of Feynman and Hibbs [\FH] and Schulman [\SCHU], and
more recently by Inomata and Singh [\INOSb], and Janke and Kleinert
[\JAKL], and second by an attempt to build in general boundary
conditions into the path integral, see e.g.\ Clark et al.\ [\CMS], and
Carreau et al.\ [\CAR, \CFG]. The best known examples are that of
bounded, but otherwise free motion, i.e.\ motion in half-spaces,
euclidean and radial potential wells, which are all standard examples.

The explicit construction of the propagator for boundary problems with,
say, Dirichlet boundary-conditions is generally quite involved. For
simple systems, however, the mirror ``principle'' can be applied.
Let us consider the boundary  at $x=0$ as totally reflecting, and denote
the reflecting action by the operator $\gamma$. The propagator
corresponding to a quantum Hamiltonian $H_0$ ($x\in\bbbr$)
\plus
$$K_0(x'',x';t'',t')=
  \Big<x''\Big\vert\e^{-\i H_0(t''-t')/\hbar}\Big\vert x'\Big>
  \Theta(t''-t')
  \tag\NUM.\num$$
of the time-evolution equation
\plus
$$\Psi(x'',t'')=\int_{\bbbr}K_0(x'',x';t'',t')\Psi(x',t')dx'
  \tag\NUM.\num$$
then can be used to construct the propagator in the half-space
$\bbbr^+$ in the following way:
\plus
$$K^{\bbbr^+}(x'',x';T)=K_0(x'',x';T)-K_0(x'',\gamma x';T)\enspace.
  \tag\NUM.\num$$
Here it must be provided that $K_0(T)$ has the invariance property
$K_0(\gamma x'', \gamma x';T)= K_0(x'',x';T)$, respectively the
corresponding Hamiltonian is invariant under the action of $\gamma$:
$\gamma H_0=H_0$. Examples, in the case of the half-space $\bbbr^+$,
are the free particle, the harmonic oscillator, or the potential
$V(x)=V_0/ \cosh^2x$ [\GROc, \GROh].

For a (free) particle in a box, there is an infinite number of
reflections and translations which must be summed over, which gives the
propagator for a particle in a box in terms of Jacobi
$\Theta$-functions, a well-known result [\FH, \SCHU].
That is, the propagator $K(T)$ is then constructed as
\plus
$$K(x'',x';T)=\sum_{\gamma\in\Gamma}K_0(x'',\gamma x';T)\enspace,
  \tag\NUM.\num$$
\edef\numaa{\NUM.\num}%
and $\Gamma$ denotes the set of all actions $\gamma$. This general
``mirror'' principle which also includes group actions plays an
important r\^ole in the theory of semiclassical periodic orbit theory
and quantum chaos (see e.g.\ [\GROu, \SIEB, \STEa] and references
therein).

However, the conceptual simplicity of Eq.~(\numaa) may be striking,
but in most cases, $K(T)$  cannot be explicitly evaluated, or  even
Eq.~(\numaa) cannot be applied because $H_0$ and $K_0(T)$ are not
invariant with respect to $\gamma$.

In this paper I want to consider quantum mechanical problems with
boundary-conditions for a wide class of systems. I will consider only
Dirichlet boundary-conditions and do not discuss generalizations
of more general local boundary-conditions
along the lines of Refs.~[\CAR, \CFG, \CMS]. The idea starts by first
considering a $\delta$-function perturbation in the path integral
[\GROh]. The path integral can be expanded into a perturbation series,
the $\delta$ functions allow to integrate over all intermediate
coordinate positions, and the convolution theorem of the
Fourier transformation decouples the time-convolutions,
giving a geometric power series in the perturbation which can be
exactly summed, yielding finally a closed expression for the
(energy-dependent) Green function of the perturbed problem (c.f.\ the
monograph by Albeverio et al.\ [\AGHH], as the main reference for a
sound mathematical definition of such $\delta$-function perturbations
in quantum mechanics).

Of course, by repeating this procedure, an arbitrary number of
$\delta$-function perturbations can be taken into account.

Dirichlet boundary-conditions at the location of the $\delta$-function
perturbation are now generated by making the strength of the
perturbation infinitely repulsive [\CMS, \MCGHU], thus yielding a
closed expression for the Green function of the boundary problem.
Having introduced the first boundary, we obtain by repeating this
procedure with a second $\delta$-function perturbation the Green
function in a box with Dirichlet boundary-conditions on both boundaries.

As I will show, it is possible to generalize  this formalism to
higher dimensions. Here an appropriate combination of $\delta$-function
perturbations along perpendicular lines, planes and hyperplanes can be
introduced, which produce by the same limiting procedure infinitely
repulsive walls at the location of the $\delta$-function perturbations.
It is obvious that arbitrarily shaped boundaries {\it cannot} be taken
into account by our formalism.

The further contents of this paper is as follows.

In the next section I shortly sketch the theory of incorporating
$\delta$-function perturbations in the path integral. The whole
theory was already presented in some length in Ref.~[\GROh] and is
not repeated here in all details. However, I will develop the
formalism further and will discuss
\medskip
\item{1)} Multiple $\delta$-function perturbations. Closed formul\ae\
          for the Green function for such arrangements were were not
          presented in [\GROh] (compare also Ref.~[\FARI], however with
          the restriction to the wave-functions only),
\item{2)} $\delta$-function perturbations along perpendicular
          lines and planes, and
\item{3)} Moving $\delta$-function perturbations.
\medskip

These results are used in the third section to construct by a limiting
procedure Green functions for boundary problems with Dirichlet
boundary-conditions. I consider half-lines, one-dimensional boxes,
half-spaces and two-dimensional boxes (with obvious generalizations
to higher dimensions), and moving boundaries.

In the fourth section I present several examples to illustrate the
formalism. The examples will include
\medskip
\item{1)} Boundary problems in half-spaces,
\item{2)} Boundary problems in boxes,
\item{3)} Boundary problems in radial boxes,
\item{4)} Boundary problems in radial rings, and
\item{5)} Moving Boundaries.

\medskip\noindent
Numerous examples for $\delta$-function perturbations in the path
integral were already given in Ref.~[\GROh] and are not repeated here.

The fifth section is devoted to a summary and discussion of the results;
in Appendix A, the derivation of the Green function for multiple
$\delta$-function perturbations is given, and in Appendix B the Green
function for the linear potential is constructed.

%----------------------------------------------------------------------
%                          END OF FILE1
%----------------------------------------------------------------------

\newpage
\glno=0                      %II
\PLUS
\section{Path Integrals with $\delta$-Function Perturbations}
\ssf
\subsection{Summation of the Perturbation Expansion}
The general method for the time-ordered perturbation expansion is quite
simple. We assume that we have a potential $W(x)=V(x)+\widetilde V(x)$ in
the path integral, where it is assumed that $W$ is so complicated
that a  direct path integration is not possible. However, the path
integral corresponding to $V(x)$ is assumed to be known. We expand the
path integral containing $\widetilde V(x)$ in a perturbation expansion about
$V(x)$ in the following way. The initial kernel corresponding to
$V$ propagates in $\Delta t$-time unperturbed, then it is interacting
with $\widetilde V$, propagates again in another $\Delta t$-time
unperturbed, a.s.o, up to the final state. This gives the series
expansion [\FH] ($x\in\bbbr$)
\plus
$$\myalign
  K(x'',x';T)
       &=\int\limits_{x(t')=x'}^{x(t'')=x''}\CD x(t)\exp\bigg\{\ih
  \int_{t'}^{t''}\bigg[{m\over2}\dot x^2-V(x)-\widetilde V(x)\bigg]dt\bigg\}
  \\   &
  =K^{(V)}(x'',x';T)+\sum_{n=1}^\infty\bigg(-\ih\bigg)^n
  {1\over n!}\left(\prod_{j=1}^n\int_{-\infty}^\infty dx_j
  \int_{t'}^{t''} dt_j\right)
  \\   &\qquad\times
  \vphantom{\bigg]^{1/2}}K^{(V)}(x_1,x';t_1-t')\widetilde V(x_1)
  K^{(V)}(x_2,x_1;t_2-t_1)
  \times\dots
  \\   &\qquad\dots\times
  \widetilde V(x_{n-1})K^{(V)}(x_n,x_{n-1};t_n-t_{n-1})
  \times\dotsc
  \\   &\qquad\dots\times
  \widetilde V(x_n)K^{(V)}(x'',x_n;t''-t_n)
  \vphantom{\bigg]^{1/2}}
  \\   &
  =K^{(V)}(x'',x';T)+\sum_{n=1}^\infty\bigg(-\ih\bigg)^n
  \left(\prod_{j=1}^n
  \int_{t'}^{t_{j+1}} dt_j\int_{-\infty}^\infty dx_j\right)
  \\   &\qquad\times
  \vphantom{\bigg]^{1/2}}K^{(V)}(x_1,x';t_1-t')
  \widetilde V(x_1)K^{(V)}(x_2,x_1;t_2-t_1)
  \times\dotsc
  \\   &\qquad\dots\times
  \widetilde V(x_{n-1})K^{(V)}(x_n,x_{n-1};t_n-t_{n-1})
  \times\dotsc
  \\   &\qquad\dots\times
  \widetilde V(x_n)K^{(V)}(x'',x_n;t''-t_n)\enspace.
  \vphantom{\bigg]^{1/2}}
  \tag\NUM.\num\endalign$$
In the second step I have ordered the time as $t'=t_0<t_1<t_2<\dots<
t_{n+1}=t''$ and paid attention to the fact that $K(t_j-t_{j-1})$ is
different from zero only if $t_j>t_{j-1}$, ([\GROh], see also e.g.\
Bauch [\BAU], Lawande and Bhagwat [\LABH]), and we set $T=t''-t'$.

We consider now an arbitrary potential $V(x)$ in one dimension with an
additional $\delta$-function perturbation so that [\GROh]
\plus
$$W(x)=V(x)-\gamma\delta(x-a)\enspace.
  \tag\NUM.\num$$
The path integral for this potential problem reads
\plus
$$\Kd(x'',x';T)
  =\int\limits_{x(t')=x'}^{x(t'')=x''}\CD x(t)\exp\bigg\{\ih
   \int_{t'}^{t''}\bigg[{m\over2}\dot x^2-V(x)+\gamma\delta(x-a)
   \bigg]dt\bigg\}\enspace.
  \tag\NUM.\num$$
We have assumed that the path integral (Feynman kernel, respectively)
for the potential $V$ is known, i.e.\ we set
\plus
$$K^{(V)}(x'',x';T)=\int\limits_{x(t')=x'}^{x(t'')=x''}
  \CD x(t)\exp\bigg\{\ih\int_{t'}^{t''}
  \bigg[{m\over2}\dot x^2-V(x)\bigg]dt\bigg\}\enspace,
  \tag\NUM.\num$$
including, of course, the (energy-dependent) Green function
\plus
$$\aligned
  \GV(x'',x';E)
 &=\ih                    \int_0^\infty dT\,\e^{\i ET/\hbar}
   K^{(V)}(x'',x';T)
  \\
  K^{(V)}(x'',x';T)&=\int_{-\infty}^{\infty}{dE\over2\pi\i}
  \e^{-\i ET/\hbar}\GV(x'',x';E)\enspace.
  \endaligned
  \tag\NUM.\num$$
\edef\numba{\NUM.\num}%
Introducing the Green function $\Gd(E)$ of the perturbed system
similarly to (\numba), it is easy to sum up the emerging geometric
power series, and we obtain due to the convolution theorem of the
Fourier transformation
\plus
$$\Gd(x'',x';E)
  =\GV(x'',x';E)-\dfrac{\GV(x'',a;E)\GV(a,x';E)}
              {\GV(a, a;E)-1/\gamma}\enspace,
  \tag\NUM.\num$$
\edef\numbb{\NUM.\num}%
where it is assumed that $\GV(a,a;E)$ actually exists. The
energy levels $E_n$ of the perturbed problem $W(x)$ are therefore
determined in a unique way by the equation
\plus
$$\GV(a,a;E_n)={1\over\gamma}\enspace.
  \tag\NUM.\num$$
The wave-functions are given by the residua of $G(E)$ at $E=E_n$, i.e.
\plus
$$\Psi_n(x)=\lim_{E\to E_n}\left[
  \dfrac{E_n-E}{1/\gamma-\GV(a,a;E)}\right]^{1/2}
  \GV(x,a;E)
  \enspace.
  \tag\NUM.\num$$
Here and in the following the index $n$ denotes the level index of a
corresponding energy level $E_n$, with $n=0,1,\dots,N_M$, and $N_M$ may
be finite or infinite.

Let us note that an implicit equation for the time-dependent propagator
is due to Gaveau and Schulman [\GASCH]. They obtained
\plus
$$\Kd(x'',x';T)=K^{(V)}(x'',x';T)+\i{\gamma\over\hbar}
  \int_{t'}^{t''}K^{(V)}(x'',a;t)\Kd(a,x';T-t)dt\enspace.
  \tag\NUM.\num$$

Similarly as for one-dimensional problems, we can consider in $D$
dimensions a radial potential according to
\plus
$$W(r)=V(r)-\gamma\delta(r-a)\enspace,
  \tag\NUM.\num$$
i.e.\ a spherical shaped $\delta$-function (also called ``shell''-,
``surface''- $\delta$-interaction [\AGHH, \AGS]) perturbation of the
potential $V(r)$. We must require $a\not=0$ because a point perturbation
leads to the evaluation of Green functions, where both arguments are
equal (and zero), an expression which in general does not exist. Of
course, we are using the usual $D$-dimensional polar coordinates. For a
radial problem we can separate variables in the path integral:
\plus
$$\myalign
  \KW(x'',x';T)&=\int\limits_{x(t')=x'}^{x(t'')=x''}\CD x(t)
  \exp\bigg\{\ih\int_{t'}^{t''}
  \bigg[{m\over2}\dot x^2-W(r)\bigg]dt\bigg\}
  \\
  &=\sum_{l=0}^\infty \KW_l(r'',r';T)
  S_l^\mu(\Omega'') S_l^{\mu*}(\Omega')\enspace,
  \tag\NUM.\num\endalign$$
where $S_l^\mu(\Omega)$ are the real hyper-spherical harmonics of
degree $l$ with unit-vector $\Omega$ [\EMOTa] and the radial path integral
$\KW_l(T)$ is given by
\plus
$$\myalign
  \KW_l&(r'',r';T)
  \\   &
  =(r'r'')^{1-D\over2}\int\limits_{r(t')=r'}^{r(t'')=r''}
  \CD r(t)\exp\bigg\{\ih\int_{t'}^{t''}
  \bigg[{m\over2}\dot r^2-\hbar^2{l(l+D-2)\over2mr^2}
                         -W(r)\bigg]dt\bigg\}
  \\   &
  \equiv(r'r'')^{1-D\over2}\int\limits_{r(t')=r'}^{r(t'')=r''}
  \mu_{l+{D-2\over2}}[r^2]\CD r(t)\exp\bigg\{\ih\int_{t'}^{t''}
  \bigg[{m\over2}\dot r^2-W(r)\bigg]dt\bigg\}\enspace.
  \tag\NUM.\num\endalign$$
\edef\numbm{\NUM.\num}%
The functional weight $\mu_\nu[r^2]$ is defined by by [\GRSb, \STEb]:
\plus
$$\align
  \mu_\nu[r^2]
  &=\lim_{N\to\infty}\prod_{j=1}^N\mu_\nu[r_{j-1}r_j]
  \tag\NUM.\num a\\
  &=\lim_{N\to\infty}\prod_{j=1}^N
  \bigg({2\pi m r_{j-1}r_j\over\i\epsilon\hbar}\bigg)^{1/2}
  \exp\bigg(-{mr_{j-1}r_j\over\i\epsilon\hbar}\bigg)
  I_\nu\bigg({mr_{j-1}r_j\over\i\epsilon\hbar}\bigg)\enspace,
  \\   &\
  \tag\NUM.\num b
  \endalign$$
in order to guarantee a well-behaved boundary behaviour at initial
at final points. $I_\nu$ describes a modified Bessel function.
Proceeding similar to the previous section and expanding the
perturbed problem into a perturbation series yields:
\plus
$$\myalign
   \ih                    \int_0^\infty dT\,\e^{\i ET/\hbar}
  &\int\limits_{r(t')=r'}^{r(t'')=r''}
  \mu_{l+{D-2\over2}}[r^2]\CD r(t)\exp\bigg\{\ih\int_{t'}^{t''}
  \bigg[{m\over2}\dot x^2-V(r)+\gamma\delta(r-a)\bigg]dt\bigg\}
  \\   &
  =\sum_{l=0}^\infty
  \Gd_l(r'',r';E) S^\mu_l(\Omega'') S^{\mu\,*}_l(\Omega')\enspace,
  \tag\NUM.\num\endalign$$
where the radial Green function is given by
\plus
$$\Gd_l(r'',r';E)=\GV_l(r'',r';E)
  -\dfrac{\GV_l(r'',a;E)\GV_l(a,r';E)}
  {\GV_l(a,a;E)-1/a^{D-1}\gamma}\enspace.
  \tag\NUM.\num$$
Therefore the energy-levels $E_n$ are determined by the equation
\plus
$${1\over a^{D-1}\gamma}=\GV_l(a,a;E_n)\enspace.
  \tag\NUM.\num$$
The corresponding wave-functions are given in a similar way as in the
one-dimensional case.

\subsection{Multiple $\delta$-Function Perturbations}
Starting from Eq.~(\numbb) we first consider the problem of two
$\delta$-function perturbations (such problems play e.g.\ a r\^ole in
the study of the Casimir effect between to perfectly conducting planes
[\BHR])
\plus
$$\multline
  \!\!\!\!
  K^{(\delta_2)}(x'',x';T)
  \\   \hfill
  =\int\limits_{x(t')=x'}^{x(t'')=x''}\CD x(t)\exp\bigg\{\ih
   \int_{t'}^{t''}\bigg[{m\over2}\dot x^2-V(x)
   +\gamma_1\delta(x-a_1)+\gamma_2\delta(x-a_2)\bigg]dt\bigg\}\enspace.
  \\ \ \endmultline
  \tag\NUM.\num$$
We obtain
\plus
$$G^{(\delta_2)}(x'',x';E)=G^{(\delta_1)}(x'',x';E)
  -\dfrac{G^{(\delta_1)}(x'',a_2;E)G^{(\delta_1)}(a_2,x';E)}
   {G^{(\delta_1)}(a_2,a_2;E)-\dfrac1{\gamma_2}}\enspace,
  \tag\NUM.\num$$
where $G^{(\delta_1)}(E)$ in turn is given by
\plus
$$G^{(\delta_1)}(x'',x';E)=\GV(x'',x';E)
  -\dfrac{\GV(x'',a_1;E)\GV(a_1,x';E)}
              {\GV(a_1,a_1;E)-\dfrac1{\gamma_1}}\enspace.
  \tag\NUM.\num$$
Insertion of $G^{(\delta_1)}(E)$ into $G^{(\delta_2)}(E)$ gives
\plus
$$\myalign
       &G^{(\delta_2)}(x'',x';E)=\GV(x'',x';E)
  \\   &
  +\bigg[\bigg(\GV(a_2,a_2;E)-{1\over\gamma_2}\bigg)
         \bigg(\GV(a_1,a_1;E)-{1\over\gamma_1}\bigg)
  \\   &\qquad\qquad\qquad\qquad
        \qquad\qquad\qquad\qquad
        -\GV(a_2,a_1;E)\GV(a_1,a_2;E)\bigg]^{-1}
  \\   &\qquad\times
  \Bigg\{\GV(a_1,x';E)\bigg[
        \GV(x'',a_2;E)\GV(a_2,a_1;E)
  \\   &\qquad\qquad\qquad\qquad
        \qquad\qquad\qquad\qquad
       -\bigg(\GV(a_2,a_2;E)-{1\over\gamma_2}\bigg)
        \GV(x'',a_1;E)\bigg]
  \\   &\qquad\quad
       +\GV(a_2,x';E)\bigg[
        \GV(x'',a_1;E)\GV(a_1,a_2;E)
  \\   &\qquad\qquad\qquad\qquad
        \qquad\qquad\qquad\qquad
       -\bigg(\GV(a_1,a_1;E)-{1\over\gamma_1}\bigg)
        \GV(x'',a_2;E)\bigg]\Bigg\}
  \\   &
  =\dfrac{\left\vert\matrix
  \pha  &               &                  \\
  \GV(x'',x';E)  &\GV(x'',a_1;E) &\GV(x'',a_2;E)    \\
  \GV(a_1,x';E)  &\GV(a_1,a_1;E)-\dfrac1{\gamma_1}
                                 &\GV(a_1,a_2;E)    \\
  \GV(a_2,x';E)  &\GV(a_2,a_1;E) &\GV(a_2,a_2;E)-\dfrac1{\gamma_2}\\
  \pha  &               &
  \endmatrix\right\vert}{\left\vert\matrix
  \pha  &                  \\
  \GV(a_1,a_1;E)-\dfrac1{\gamma_1}
                 &\GV(a_1,a_2;E)    \\
  \GV(a_2,a_1;E) &\GV(a_2,a_2;E)-\dfrac1{\gamma_2}\\
  \pha  &                  \endmatrix\right\vert}\enspace.
  \tag\NUM.\num\endalign$$
\edef\numbc{\NUM.\num}%
Bound states are determined by
\plus
$$\left\vert\matrix
  \GV(a_1,a_1;E_n)-\dfrac1{\gamma_1}
                   &\GV(a_1,a_2;E_n)    \\
  \GV(a_2,a_1;E_n) &\GV(a_2,a_2;E_n)-\dfrac1{\gamma_2}
  \endmatrix\right\vert=0\enspace.
  \tag\NUM.\num$$
The bound state wave-functions are given by
\plus
$$\Psi_n^*(x')\Psi_n(x'')
  =\lim_{E\to E_n}\Big[(E_n-E)G^{(\delta_2)}(x'',x';E)\Big]
  \enspace.
  \tag\NUM.\num$$
Eq.~(\numbc) has an obvious generalization to the problem of
$N$ $\delta$-functions.

\eject\noindent
We obtain
\plus
$$\myalign
  &G^{(\delta_N)}(x'',x';E)
  \\   &
  =\ih                    \int_0^\infty dT\,\e^{\i ET/\hbar}
  \int\limits_{x(t')=x'}^{x(t'')=x''}\CD x(t)\exp\left\{\ih
   \int_{t'}^{t''}\left[{m\over2}\dot x^2-V(x)
   +\sum_{k=1}^N\gamma_k\delta(x-a_k)\right]dt\right\}
  \\   &
  =\dfrac{\left\vert\matrix
  \pha  &               &                  &        \\
  \GV(x'',x';E)  &\GV(x'',a_1;E) &\hdots
                                 &\GV(x'',a_N;E)             \\
  \GV(a_1,x';E)  &\GV(a_1,a_1;E)-\dfrac1{\gamma_1}
                 &\hdots         &\GV(a_1,a_N;E)             \\
  \vdots         &\vdots         &\ddots            &\vdots  \\
  \GV(a_N,x';E)  &\GV(a_N,a_1)   &\hdots
                                 &\GV(a_N,a_N;E)-\dfrac1{\gamma_N}\\
  \pha  &               &                  &
  \endmatrix\right\vert}{\left\vert\matrix
  \pha  &               &                  \\
  \GV(a_1,a_1;E)-\dfrac1{\gamma_1}
                 &\hdots         &\GV(a_1,a_N;E)    \\
  \vdots         &\ddots         &\vdots            \\
  \GV(a_N,a_1;E) &\hdots         &\GV(a_N,a_N;E)-\dfrac1{\gamma_N}\\
  \pha  &               &   \endmatrix\right\vert}\enspace.
  \tag\NUM.\num\endalign$$
\edef\numbd{\NUM.\num}%
The quantization condition for the of bound states has the form
\plus
$$\left\vert\matrix
  \GV(a_1,a_1;E_n)-\dfrac1{\gamma_1}
                   &\hdots    &\GV(a_1,a_N;E_n)    \\
  \vdots           &\ddots    &\vdots              \\
  \GV(a_N,a_1;E_n) &\hdots    &\GV(a_N,a_N;E_n)-\dfrac1{\gamma_N}
  \endmatrix\right\vert=0\enspace.
  \tag\NUM.\num$$
This can be written for short by
\plus
$$\det(a_{ij})=0\enspace,\qquad
   \hbox{with the $N\times N$ matrix}\qquad
   (a_{ij})
   =\bigg(\GV(a_i,a_j;E)-{\delta_{ij}\over\gamma_i}\bigg)\enspace.
   \tag\NUM.\num$$
The bound state wave-functions are given by
\plus
$$\Psi_n^*(x')\Psi_n(x'')
  =\lim_{E\to E_n}\Big[(E_n-E)G^{(\delta_N)}(x'',x';E)\Big]
  \enspace.
  \tag\NUM.\num$$
Eq.~(\numbd) is proven in appendix A. Let us note that in the limit
$N\to\infty$ Eq.(\numbd) can be used to describe periodic
$\delta$-functions in solid state physics, see e.g.\ Goovaerts and
Broeckx [\GOBR] for such a discussion, including the determination of
energy bands.

\eject
\subsection{$\delta$-Function Perturbation Along a Line}
The case of an arbitrary located two-dimensional $\delta$-function
perturbation on a line can now also be treated. We consider a
$\delta$-function perturbation $\propto-\gamma\delta(ax-by-c)$. This
can be alternatively interpreted as two particles interacting pointwise
with coupling $\gamma$. By a simple shift of variables and introducing
center of mass and relative coordinates $R$ and $r$, respectively
$$\alignat 3
    r&=ax-by-c\enspace,
     &\qquad
  \mu&={m\over a^2+b^2}\enspace,
  \tag\NUM.\num\\   \global\plus
    R&={ab^2x+a^2(by+c)\over a^2+b^2}\enspace,
     &\qquad
    M&=m{a^2+b^2\over a^2b^2}\enspace,
  \tag\NUM.\num\endalignat$$
we obtain by using the known exact propagator for the $\delta$-function
perturbation (e.g.\ [\GROh] and references therein) the path integral
identity
\plus
$$\myalign
       &
  \int\limits_{x(t')=x'}^{x(t'')=x''}\CD x(t)
  \int\limits_{y(t')=y'}^{y(t'')=y''}\CD x(t)
  \exp\left\{\ih\int_{t'}^{t''}\bigg[{m\over2}
     (\dot x^2+\dot y^2)+\gamma\delta(ax-by-c)\bigg]dt\right\}
  \\   &
  =ab\int\limits_{R(t')=R'}^{R(t'')=R''}\CD R(t)
  \exp\left({\i M\over2\hbar}\int_{t'}^{t''}\dot R^2\right)
  \\   &\qquad\times
  \int\limits_{r(t')=r'}^{r(t'')=r''}\CD r(t)
  \exp\left\{\ih\int_{t'}^{t''}\bigg[{\mu\over2}
     \dot r^2+\gamma\delta(r)\bigg]dt\right\}
  \\   &
  =\sqrt{M\over2\pi\i\hbar T}
          \exp\bigg[-{M\over2\hbar\i T}(R''-R')^2\bigg]
  \\   &\qquad\times
   \Bigg\{
   {\mu\gamma\over\hbar^2}\exp\bigg[-{\mu\gamma\over\hbar^2}(r'+r'')
          +\ih{\mu\gamma^2\over2\hbar^2}T\bigg]
  +{1\over2\pi}\int_{-\infty}^\infty dp
  \exp\bigg(-\i{p^2\hbar T\over2\mu}\bigg)
  \\   &\qquad\qquad\qquad\qquad\times
  \Bigg[\sin pr'\sin pr''+\cos pr'\cos pr''
  -{\e^{\i p(r'+r'')}\over1+\i{p\hbar^2\over\mu\gamma}}\Bigg]\Bigg\}
  \enspace.
  \tag\NUM.\num\endalign$$
There seems no obvious way to incorporate more than one
arbitrary line in $\bbbr^2$ equipped with $\delta$-function
perturbations. However, the two-dimensional result can be generalized
to $D$ dimensions. We introduce a $\delta$-function perturbation on a
hyper-plane $\vec a\cdot\vec x-b=0$ ($\vec a, \vec x\in\bbbr^D$).
Introducing appropriate center of mass and relative coordinates [\MCG,
\MCGHU], we obtain similarly as in the two-dimensional case a
$(D-1)$-dimensional product of free particle kernels and one
$\delta$-function potential kernel in the relative coordinate.

\subsection{$\delta$-Function Perturbations along Perpendicular Lines
            and Planes}
Let us consider first a two-dimensional system. Obviously we have the
identity
\plus
$$\myalign
  &\int\limits_{x(t')=x'}^{x(t'')=x''}\CD x(t)
   \int\limits_{y(t')=y'}^{y(t'')=y''}\CD y(t)
   \exp\Bigg\{\ih
   \int_{t'}^{t''}\Bigg[{m\over2}(\dot x^2+\dot y^2)-V(x)-V(y)
   \\  &\qquad\qquad\qquad\qquad
   +\sum_{k=1}^{N_1}\gamma_{1,k}\delta(x-a_k)
   +\sum_{k=1}^{N_2}\gamma_{2,k}\delta(y-b_k)\Bigg]dt\Bigg\}
  \\   &
  =\int\limits_{x(t')=x'}^{x(t'')=x''}\CD x(t)\exp\left\{\ih
   \int_{t'}^{t''}\left[{m\over2}\dot x^2-V(x)
   +\sum_{k=1}^{N_1}\gamma_{1,k}\delta(x-a_k)\right]dt\right\}
  \\   &\qquad\times
   \int\limits_{y(t')=y'}^{y(t'')=y''}\CD y(t)\exp\left\{\ih
   \int_{t'}^{t''}\left[{m\over2}\dot y^2-V(y)
   +\sum_{k=1}^{N_2}\gamma_{2,k}\delta(y-b_k)\right]dt\right\}
  \\   &
  \equiv K^{(N_1)}_x(x'',x';T)\cdot K^{(N_2)}_y(y'',y':T)
  =K^{(N_1,N_2)}(x'',x',y'',y';T)
  \enspace.
  \tag\NUM.\num\endalign$$
\edef\numbn{\NUM.\num}%
The solution of each kernel in terms of its corresponding
Green function is given by Eq.~(\numbd). Now call these Green
functions $G^{(N_1)}_x(E)$ and $G^{(N_2)}_y(E)$, respectively. The
convolution theorem of the Fourier transformation now states
that the Green function corresponding to the entire kernel
$G^{(N_1,N_2)}(x'',x',y'',y';E)$ is given by the convolution of
$G^{(N_1)}_x(E)$ and $G^{(N_2)}_y(E)$, i.e.
\plus
$$G^{(N_1,N_2)}(x'',x',y'',y';E)=\int_{-\infty}^\infty dz\,
  G^{(N_1)}_x(x'',x';z)G^{(N_2)}_y(y'',y';E-z)  \enspace.
  \tag\NUM.\num$$
Obviously, we can generalize this result to the $D$-dimensional case.
By separating the $D$ path integrations we obtain a $D$-fold product of
one-dimensional multiple $\delta$-function potential kernels. The Green
function of the entire kernel is given by a $(D-1)$-fold convolution of
the corresponding one-dimensional multiple $\delta$-function potential
Green function, each of them given by Eq.~(\numbd). As complicated as
this may look, for practical purposes only the case $D=3$ is relevant,
and what remains is a two-fold convolution. Because the number of
$\delta$-functions we take into account is not restricted, nor their
variability in strength, we can take this model in the large $N_1,N_2$
limit of perpendicular $\delta$-function perturbations as a model for a
rigid body lattice. The various entries in the determinants are simple,
and all what one has to do, is the evaluation of the determinant and
some numerical integration. Of course, the whole system can also be put
into a ($D$-dimensional) box (see next section) which only slightly
increases the complexity of the system. A numerical investigation
should yield a band structure of the energy levels.

\eject
\subsection{Moving $\delta$-Function Perturbation}
Explicitly time-dependent problems in quantum mechanics turn out to be
usually quite difficult and involved. In concrete physical situations
one encounters mostly scattering problems and it is necessary to study
the quantum mechanical properties of a system with a moving background,
e.g.\ electrons scattered by moving ions. One useful model of such
set-ups are $\delta$-function perturbations, simulating point
interactions (e.g.\ [\KOM, \ZHCH]). In this Subsection we study the
quantum motion of a particle in the field of a moving $\delta$-function
perturbation [\ZHCH]:
\plus
$$V(x,t)=-\gamma\delta(x-vt)\enspace,
  \tag\NUM.\num$$
where $v$ denotes the constant velocity of the $\delta$-function.
I present an alternative treatment as given by Duru [\DURf] (compare
also [\GROu]) based directly on the summation of the corresponding
perturbation expansion. The path integral now has the form
\plus
$$\myalign
  &\Kd(x'',x';t'',t')
  \\   &=
  \int\limits_{x(t')=x'}^{x(t'')=x''}\CD x(t)
  \exp\left\{\ih\int_{t'}^{t''}
   \bigg[{m\over2}\dot x^2+\gamma\delta(x-vt)\bigg]dt\right\}
  \\   &=
  \lim_{N\to\infty}\bigg({m\over2\pi\i\epsilon\hbar}\bigg)^{N/2}
  \prod_{j=1}^{N-1}\int_{-\infty}^\infty dx_j
  \exp\left\{\ih\sum_{j=1}^N\bigg[{m\over2\epsilon}(x_j-x_{j-1})^2
        +\epsilon\gamma\delta(x_j-vt_j)\bigg]\right\}\enspace,
  \\  &
  \tag\NUM.\num\endalign$$
($x_j=x(t_j)$, $t_j=t'+\epsilon j$, with $\epsilon=(t''-t')/N=T/N$ in
the limit $N\to\infty$). We first perform the coordinate substitution
\plus
$$y=x-vt\enspace,\qquad\hbox{with}\qquad
  \left\{\aligned  y''&=x''-vt''\enspace,\\  y'&=x'-vt'\enspace.
  \endaligned\right.
  \tag\NUM.\num$$
This yields
\plus
\hfuzz=4pt
$$\myalign
  &\Kd(x'',x';t'',t')
  \\  &=
  \exp\bigg(\ih{mv^2 T\over2}\bigg)
  \int\limits_{y(t')=y'}^{y(t'')=y''}\CD y(t)
  \exp\left\{\ih\int_{t'}^{t''}
  \bigg[{m\over2}\dot y^2+mv\dot y+\gamma\delta(y)\bigg]\right\}
  \Bigg\vert_{{y'=x'-vt'\hfill\atop y''=x''-vt''}},
  \\  &=
  \sqrt{m\over2\pi\i\hbar T}\,\exp\bigg[
  -{m\over2\i\hbar T}(x''-x')^2\bigg]
  +\exp\bigg(\ih{mv^2T\over2}\bigg)
  {\widetilde K}^{(\delta)}(y'',y';t'',t')
  \Bigg\vert_{{y'=x'-vt'\hfill\atop y''=x''-vt''}} ,
  \\  &
  \tag\NUM.\num\endalign$$
\hfuzz=3pt
\eject\noindent
where ${\widetilde K}^{(\delta)}$ is given by
\plus
$$\myalign
  &{\widetilde K}^{(\delta)}(y'',y';t'',t')
  \\   &\qquad
  =\sum_{n=1}^\infty\bigg({\i\gamma\over\hbar}\bigg)^n
  \int_{t'}^{t''}dt_n\dots\int_{t'}^{t_2}dt_1
  \prod_{j=1}^n\int_{-\infty}^\infty dy_j\delta(y_j)
  \\   &\qquad\qquad\times
  K^{(v)}(y_1,y';t',t_1)
  K^{(v)}(y_2,y_1;t_2,t_1)
  \dots
  K^{(v)}(y'',y_n;t'',t_n)
  \\    &\qquad
  =\sum_{n=1}^\infty\bigg({\i\gamma\over\hbar}\bigg)^n
  \int_{t'}^{t''}dt_n\dots\int_{t'}^{t_2}dt_1
  \\   &\qquad\qquad\times
  K^{(v)}(0,y';t',t_1)
  \cdot\prod_{j=2}^{n-1}
  K^{(v)}(0,0;t_j,t_{j-1})
  \cdot K^{(v)}(y'',0;t'',t_n).
  \tag\NUM.\num\endalign$$
Here $\widetilde K^{(v)}(t'',t')$ denotes the path integral
\plus
$$\myalign
  \widetilde K^{(v)}(y'',y';t'',t')
  &=\int\limits_{y(t')=y'}^{y(t'')=y''}\CD y(t)
  \exp\left[\ih\int_{t'}^{t''}
     \bigg({m\over2}\dot y^2+mv\dot y\bigg)dt\right]\\
  &=\sqrt{m\over2\pi\i\hbar T}\,\exp\bigg[
   -{m\over2\i\hbar T}(y''-y')^2+\i{mv\over\hbar}(y''-y')\bigg]\enspace.
  \tag\NUM.\num\endalign$$
We find
\plus
$$G^{(v)}(y'',y';E)=\overh\sqrt{-{m\over2E}}\,
  \exp\bigg[-{\vert y''-y'\vert \over\hbar}
  \sqrt{-2mE}+\i{mv\over\hbar}(y''-y')\bigg]\enspace,
  \tag\NUM.\num$$
and we can apply the convolution theorem of the Fourier
transformation yielding
\plus
$$\widetilde \Gd(y'',y';E)
  ={m\gamma\over2\hbar^2}
  \thickfrac{\dsize\exp\left[
  -{\vert y'\vert +\vert y''\vert \over\hbar}\sqrt{-2mE}\,
             +\i{mv\over\hbar}(y''-y')\right]}
  {\dsize\sqrt{-E}\left(\sqrt{-E}
   -{\gamma\over\hbar}\sqrt{m\over2}\right)}.
  \tag\NUM.\num$$
$K(t'',t')$ can be easily calculated by inverse Laplace transformation
[\EMOTb, Vol.I.\ p.247]
\plus
$$\CL^{-1}\Big[p^{-1/2}(p^{1/2}+\beta)^{-1}\e^{-\alpha\sqrt{p}}\Big]
  (t)=
  \e^{\alpha\beta+\beta^2t}\erfc\bigg({\alpha\over2\sqrt{t}}
    +\beta\sqrt{t}\bigg)\enspace,
  \tag\NUM.\num$$
yielding finally (re-inserting $x'$ and $x''$)
\eject\noindent
\plus
$$\myalign
    \Kd&(x'',x';t'',t')
  \\   &
  =\sqrt{m\over2\pi\i\hbar T}
   \exp\bigg[{\i m\over2\hbar T}(x''-x')^2\bigg]
  \\   &\qquad
  +{m\gamma\over2\hbar^2}
  \exp\Bigg\{{\i\over\hbar}\bigg[v(x''-xt'')-v(x'-vt')+m{v^2\over2}T
      \bigg]\Bigg\}
  \\   &\qquad\qquad\times
  \exp\Bigg[-{m\gamma\over\hbar^2}(\vert x''-vt''\vert
   +\vert x'-vt'\vert)+\ih{m\gamma^2\over2\hbar}T\Bigg]
  \\   &\qquad\qquad\times
  \erfc\Bigg[\sqrt{m\over2\i\hbar T}\,
       \bigg(\vert x''-vt''\vert+\vert x'-vt'\vert
          -\ih\gamma T\bigg)\Bigg]
  \\   &
  ={m\gamma\over\hbar^2}
  \exp\bigg[
   -{m\gamma\over\hbar^2}
    \bigg(\vert x''-vt''\vert +\vert x'-vt'\vert\bigg)
  \\   &\qquad\qquad\qquad
  -{\i mv\over\hbar}(x'-vt')+{\i mv\over\hbar}(x''-vt'')
  +{\i m\over2\hbar}\bigg({\gamma^2\over\hbar^2}+v^2\bigg)T\bigg]
  \\   &
  +{1\over2\pi}\int_{-\infty}^\infty dp
  \exp\bigg(-\i{p^2\hbar T\over2m}\bigg)
  \left\{\e^{\i p(x''-x')}
  \vphantom{\dfrac{\dsize{\i mv^2\over2\hbar}T}
   {1+\i\dfrac{p\hbar^2}{m\gamma}}}
  \right.\\  &\left.\qquad\qquad\qquad
  -\thickfrac{\dsize
    \exp\left[\i p(\vert x''-vt''\vert+\vert x'-vt'\vert)
      +{\i mv\over\hbar}\bigg(x''-x'-{vT\over2}\bigg)\right]}
   {1+\i\dfrac{p\hbar^2}{m\gamma}}\right\}\enspace.
  \\   &
  \tag\NUM.\num\endalign$$
We obtain one bound state
\plus
$$\Psi^{(bound)}(x,t)={\sqrt{m\gamma}\over\hbar}
  \exp\bigg[-{m\gamma\over\hbar^2}\vert x-vt\vert
 +{\i mv\over\hbar}(x-vt)
  +{\i m\over2\hbar}\bigg({\gamma^2\over\hbar^2}+v^2\bigg)t\bigg],
  \tag\NUM.\num$$
which is the result of Ref.[\ZHCH], and the continuous states have the form
\plus
$$\multline
  \Psi^{(cont.)}(x,t)={1\over\sqrt{2\pi}}
  \exp\bigg(-{\i p^2\hbar t\over2m}\bigg)
  \\   \times
  \left\{\e^{\i px}-
  \dfrac{\exp\left[\i p\vert x-vt\vert +\dfrac{\i mv}{\hbar}(x-vt)
  +\dfrac{\i mv^2}{2\hbar}t\right]}
  {1+\i\dfrac{p\hbar^2}{m\gamma}}\right\}.
  \endmultline
  \tag\NUM.\num$$
For $v=0$ the usual result is recovered.

%----------------------------------------------------------------------
%                          END OF FILE2
%----------------------------------------------------------------------

\eject\noindent
\glno=0                      %III
\PLUS
\section{Path Integrals for Boundary Problems}\ssf
\subsection{Motion in Half-Spaces and Boxes}
In Eq.~(\numbb) we consider now the limit $\gamma\to-\infty$ which has
the effect that an impenetrable wall appears at $x=a$ [\CMS, \MCGHU].
We set $\lim_{\gamma\to-\infty}\Gd(E)\equiv G^{(Wall)}(E)$ i.e.\ we
obtain
\plus
$$\align
  \GWa&(x'',x';E)
  \\   &
  =\ih                    \int_0^\infty dT\,\e^{\i ET/\hbar}
  \int\limits_{x(t')=x'}^{x(t'')=x''}\CD_{Wall}x(t)
  \exp\left\{\ih\int_{t'}^{t''}\bigg[{m\over2}\dot x^2
       -V(x)\bigg]dt\right\}
  \\   &
  =\GV(x'',x';E)-\dfrac{\GV(x'',a;E)\GV(a,x';E)}{\GV(a, a;E)}\enspace.
  \tag\NUM.\num\endalign$$
\edef\numbh{\NUM.\num}%
Bound states are determined by the equation
\plus
$$\GV(a,a;E_n)=0\enspace.
  \tag\NUM.\num$$
Again, the wave-functions are given by the residua of $G(E)$ at
$E=E_n$, i.e.
\plus
$$\Psi_n(x)=\lim_{E\to E_n}\left[
  -\dfrac{E_n-E}{\GV(a,a;E)}\right]^{1/2}
  \GV(x,a;E)\enspace.
  \tag\NUM.\num$$
The Green functions $\GV(E)$ usually can be written as a product
of two linearly independent solutions of the corresponding
homogeneous Schr\"odinger equation, i.e.\ we can consider in general
the Ansatz for $\GV(E)$
\plus
$$\GV(x'',x';E)=f_E A_E(x_>)B_E(x_<)\enspace,
  \tag\NUM.\num$$
\edef\numbe{\NUM.\num}%
where $f_E$ is a prefactor which depend only on $E$, and
$A_E(x_>)$ and $B_E(x_<)$ are functions of the larger (smaller) of
$x',x''$, and depending, of course, on $E$ as an parameter.
We set the barrier at $x=R$ and consider the motion for
the (right-) half-space $x>R$. This yields
\plus
$$G^{(x>R)}(x'',x';E)=f_E A_E(x_>)B_E(x_<)
       -{f_EB_E(R)A_E(x')A_E(x'')\over A_E(R)}\enspace
  \tag\NUM.\num$$
\edef\numbk{\NUM.\num}%
and the quantization condition has the form
\plus
$$A_{E_n}(R)=0\enspace.
  \tag\NUM.\num$$
Similarly, if we put the barrier at $x=a$ and consider the motion
in the (left-) half-space $x<a$, we obtain
\plus
$$G^{(x<a)}(x'',x';E)=f_E A_E(x_>)B_E(x_<)
       -{f_EA_E(a)B_E(x')B_E(x'')\over B_E(a)}\enspace
  \tag\NUM.\num$$
\edef\numbl{\NUM.\num}%
and the quantization condition has the form
\plus
$$B_{E_n}(a)=0\enspace.
  \tag\NUM.\num$$

Repeating the procedure for the double $\delta$-function perturbation,
we consider the limit $\lim_{\gamma_1,\gamma_2\to-\infty}
G^{(\delta_2)}(E)\equiv G^{(Box)}(E)$ and obtain
for the motion in the box $a<x<b$
\plus
$$\align
  G^{(Box)}&(x'',x';E)
  \\   &
  =\ih                    \int_0^\infty dT\,\e^{\i ET/\hbar}
  \int\limits_{x(t')=x'}^{x(t'')=x''}\CD_{Box}x(t)
  \exp\left\{\ih\int_{t'}^{t''}\bigg[{m\over2}\dot x^2
       -V(x)\bigg]dt\right\}
  \\   &
  ={\left\vert
  \matrix \GV(x'',x';E) &\GV(x'',b;E) &\GV(x'',a;E)  \\
          \GV(b,x';E)   &\GV(b,b;E)   &\GV(b,a;E)    \\
          \GV(a,x';E)   &\GV(a,b;E)   &\GV(a,a;E)
  \endmatrix\right\vert\over\left\vert\matrix
                       &              \\
          \GV(b,b;E)   &\GV(b,a;E)    \\
          \GV(a,b;E)   &\GV(a,a;E)    \\
                       &          \endmatrix\right\vert}
  \enspace.
  \tag\NUM.\num\endalign$$
\edef\numbf{\NUM.\num}%
The quantization condition for the (infinite number of) bound states
has the form
\plus
$$\left\vert\matrix
  \GV(b,b;E_n)   &\GV(b,a;E_n)    \\
  \GV(a,b;E_n)   &\GV(a,a;E_n)\endmatrix\right\vert=0
  \enspace.
  \tag\NUM.\num$$
The bound state wave-functions are given by
\plus
$$\Psi_n^*(x')\Psi_n(x'')
  =\lim_{E\to E_n}\Big[(E_n-E)G^{(Box)}(x'',x';E)\Big]
  \enspace.
  \tag\NUM.\num$$
Taking the same Ansatz (\numbe) as before for the Green function
$\GV(E)$ we can simplify Eq.~(\numbf) into
\plus
$$\multline
  \GB(x'',x';E)=f_E A_E(x_>)B_E(x_<)
  +{f_E\over A_E(a)B_E(b)-A_E(b)B_E(a)}
  \\    \times
  \Bigg\{A_E(x'')B_E(a)\Big[B_E(x')A_E(b)-A_E(x')B_E(b)\Big]
  \\    +B_E(x'')A_E(b)\Big[A_E(x')B_E(a)-B_E(x')A_E(a)\Big]\Bigg\}
  \enspace,
  \endmultline
  \tag\NUM.\num$$
and the quantization condition has the form
\plus
$$A_{E_n}(a)B_{E_n}(b)-A_{E_n}(b)B_{E_n}(a)=0\enspace.
  \tag\NUM.\num$$
For radial problems with a boundary at $r=R$ we obtain similarly
\plus
$$\myalign
   \ih                    \int_0^\infty dT\,\e^{\i ET/\hbar}
  &\int\limits_{x(t')=x'}^{x(t'')=x''}\CD_{Box} x(t)
  \exp\bigg\{\ih\int_{t'}^{t''}
  \bigg[{m\over2}\dot x^2-V(r)\bigg]dt\bigg\}
  \\   &
  =\sum_{l=0}^\infty
  G_l^{(Box)}(r'',r';E) S^\mu_l(\Omega'') S^{\mu\,*}_l(\Omega')\enspace,
  \tag\NUM.\num\endalign$$
where the radial Green function is given by
\plus
$$G_l^{(Box)}(r'',r';E)=\GV_l(r'',r';E)
  -\dfrac{\GV_l(r'',R;E)\GV_l(R,r';E)}{\GV_l(R,R;E)}\enspace.
  \tag\NUM.\num$$
\edef\numbi{\NUM.\num}%
The energy levels are determined by
\plus
$$\GV_l(R,R;E_n)=0\enspace.
  \tag\NUM.\num$$
For a radial ring with motion constraint by $a<r<b$ we obtain
\plus
$$G_l^{(Ring)}(r'',r;E)
  ={\left\vert
  \matrix                 &               &                \\
          \GV_l(x'',x';E) &\GV_l(x'',b;E) &\GV_l(x'',a;E)  \\
          \GV_l(b,x';E)   &\GV_l(b,b;E)   &\GV_l(b,a;E)    \\
          \GV_l(a,x';E)   &\GV_l(a,b;E)   &\GV_l(a,a;E)    \\
                          &               &
  \endmatrix\right\vert\over\left\vert\matrix
                          &                \\
          \GV_l(b,b;E)    &\GV_l(b,a;E)    \\
          \GV_l(a,b;E)    &\GV_l(a,a;E)    \\
                          &             \endmatrix\right\vert}
  \enspace,
  \tag\NUM.\num$$
\edef\numbj{\NUM.\num}%
and the bound states energy levels are determined by
\plus
$$\left\vert\matrix
  \GV_l(b,b;E_n)    &\GV_l(b,a;E_n)    \vphantom{\sum}    \\
  \GV_l(a,b;E_n)    &\GV_l(a,a;E_n)\endmatrix\right\vert=0
  \enspace.
  \tag\NUM.\num$$
The bound state wave-functions can be obtained from
\plus
$$\Psi_{ln}^*(x')\Psi_{ln}(x'')
  =\lim_{E\to E_n}\Big[(E_n-E)G_l^{(Ring)}(x'',x';E)\Big]
  \enspace.
  \tag\NUM.\num$$

\subsection{Boundaries Along Perpendicular Lines and Planes}
Because the Green functions for problems with Dirichlet
boundary-conditions can be constructed from the corresponding Green
functions for $\delta$-function perturbations we can also discuss the
case of multidimensional boxes. We consider Eq.~(\numbn). Since we are
only interested in one domain which emerges form the limit $\gamma
_{1/2,k}\to-\infty$, it is sufficient to consider the case $N_1=N_2=2$.
According to the convolution theorem this gives the Green function for
a two-dimensional system in a box $a_1<x<b_1$, $a_2<y<b_2$
\plus
$$\align
  &\ih                    \int_0^\infty dT\,\e^{\i ET/\hbar}
  \\   &\quad\times\!\!
  \int\limits_{x(t')=x'}^{x(t'')=x''}\!\!\CD_{Box}x(t)\!\!
  \int\limits_{y(t')=y'}^{y(t'')=y''}\!\!\CD_{Box}y(t)
  \exp\left\{\ih\int_{t'}^{t''}\bigg[{m\over2}(\dot x^2+\dot y^2)
       -V(x)-V(y)\bigg]dt\right\}
  \\   &
  =\int_{-\infty}^\infty dz\,
  G^{(Box)}_x(x'',x';z)G^{(Box)}_y(y'',y';E-z)  \enspace.
  \tag\NUM.\num\endalign$$
This results can also be generalized in an obvious way to
higher dimensions with the restriction that no more perpendicular
planes are allowed than twice the dimension of the space, i.e.\
$\#\hbox{(boundaries)}\leq2D$. The entire Green function is then
given by a $(D-1)$-fold convolution of Green functions (\numbf).
$\GV(E)$ itself can include $\delta$-function perturbations, c.f.\ the
remark at the end of the last section.

\subsection{Moving Boundaries}
Actually, the problem of a moving $\delta$-function perturbation belongs
to a larger class of explicitly time-dependent potentials. One possibility
is to change a usual potential according to $V(x)\mapsto V(x/\zeta)/
\zeta^2$, where $\zeta=\zeta(t)=(at^2+2bt+c)^{1/2}$, and one derives the
path integral identity [\DMN, \GROw]
\plus
$$\multline
   \int\limits_{x(t')=x'}^{x(t'')=x''}\CD x(t)
  \exp\left\{\ih\int_{t'}^{t''}\left[{m\over2}\dot x^2
     -{1\over\zeta^2(t)}V\bigg({x\over\zeta(t)}\bigg)\right]dt\right\}
  \\
  =\big(\zeta''\zeta'\big)^{-D/2}
  \exp\left[{\i m\over2\hbar}\left({x''}^2{\dot\zeta''\over\zeta''}
                     -{x'}^2{\dot\zeta'\over\zeta'}\right)\right]
  K_{\omega',V}\bigg({x''\over\zeta''},{x'\over\zeta'};
  \tau(t'')\bigg)\enspace,
  \endmultline
  \tag\NUM.\num$$
\edef\numca{\NUM.\num}%
with $\zeta'=\zeta(t')$, $\zeta''=\zeta(t'')$, etc., and we have set
$\tau(t'')=\int_{t'}^{t''}dt/\zeta^2(t)$. Furthermore
${\omega'}^2=ac-b^2$ and $K_{\omega',V}$ denotes the path integral
\plus
$$K_{\omega',V}(z'',z';s'')
  =\int\limits_{z(0)=z'}^{z(s'')=z''}\CD z(s)
  \exp\left\{\ih\int_0^{s''}\bigg[{m\over2}\dot z^2
   -{m\over2}{\omega'}^2z^2-V(z)\bigg]ds\right\}\enspace.
  \tag\NUM.\num$$
Another class of time-dependent problems has a time-dependence
according to $V(x)\mapsto V(x-f(t))$. Here one gets [\DURf]
($q'=x'-f'$, $f'=f(t')$, etc.)
\plus
$$\multline
  \int\limits_{x(t')=x'}^{x(t'')=x''}\CD x(t)
  \exp\left\{\ih\int_{t'}^{t''}\bigg[
  {m\over2}\dot x^2-V(x-f(t))\bigg]dt\right\}
  \\   \qquad
  =\exp\left\{{\i m\over\hbar}
  \left[\dot f''(x''-f'')-\dot f'(x'-f')
  +\half\int_{t'}^{t''}\dot f^2(t)dt\right]\right\}
  K_{\ddot f,V}(q'',q';T)
  \hfill\endmultline
  \tag\NUM.\num$$
\edef\numcb{\NUM.\num}%
with the path integral $K_{\ddot f,V}(T)$ given by
\plus
$$K_{\ddot f,V}(q'',q';T)
  =\int\limits_{q(t')=q'}^{q(t'')=q''}\CD q(t)
  \exp\left\{\ih\int_{t'}^{t''}\bigg[
  {m\over2}\dot q^2-V(q)-m\ddot f(t) q\bigg]dt\right\}\enspace,
  \tag\NUM.\num$$
Hence, in both cases the {\it time-dependent\/} problem can be rewritten
in terms of a {\it time-independent\/} problem, however, additional terms
appear in the Lagrangian. Considering now time-dependent $\delta$-function
perturbations, the time dependence can be transformed away and one is left
with the path integrals $K_{\omega',\delta}(\tau(t''))$ and
$K_{\ddot f,\delta}(T)$, respectively, with the $\delta$-function
perturbation located at the coordinate origin. Let us denote the
corresponding Green functions by $G_{\omega',\delta}(E)$ and
$G_{\ddot f,\delta}(E)$. Considering the limiting process of making the
strength of the $\delta$-function perturbation infinite repulsive, gives
the corresponding Green functions $G_{\omega',Wall}(E)$ and
$G_{\ddot f,Wall}(E)$ according to Eq.~(\numbh), with the totally
reflecting boundary located at the origin, which describe still
time-independent problems. The {\it time-dependent\/} problem can be
obtained by applying Eqs.~(\numca,\numcb), provided we can explicitly
evaluate $K_{\omega',Wall}(\tau(t''))$ and $K_{\ddot f,Wall}(T)$.
It is obvious that this last step is a rather strong restriction which
allows only for a limited number of explicitly soluble examples, say if we
consider $\omega'=0$ and $\ddot f=0$, respectively. One then obtains
[(MB) -  moving boundary, $D=1$]
\plus
$$\multline
  K^{(MB)}(x'',x';t'',t')
  \\   \qquad
  =\big(\zeta''\zeta'\big)^{-1/2}
  \exp\left[{\i m\over2\hbar}\left({x''}^2{\dot\zeta''\over\zeta''}
                     -{x'}^2{\dot\zeta'\over\zeta'}\right)\right]
  K_{\omega',Wall}\bigg({x''\over\zeta''},{x'\over\zeta'};
  \tau(t'')\bigg)\enspace,
  \hfill\endmultline
  \tag\NUM.\num$$
and similarly as for the case of Eq.~(\numcb).

%----------------------------------------------------------------------
%                          END OF FILE3
%----------------------------------------------------------------------

\PLUS
\glno=0                      %IV
\section{Examples}
In this section I am going to discuss several examples of the
formalism presented in the previous section. I do not discuss examples
with $\delta$-function perturbations, this has been already done in
Ref.~[\GROh].

\subsection{Boundary Problems in Half-Space}
\subsubsection{Free Motion in Half-Space}
As the first trivial example we consider for completeness the free
motion in the half-space $x>0$ ($\bbbr^+$). We have for the free motion
in $\bbbr$
$$\myalign
  \int\limits_{x(t')=x'}^{x(t'')=x''}\CD x(t)
  \exp\left({\i m\over2\hbar}\int_{t'}^{t''}\dot x^2dt\right)
  &=\sqrt{m\over2\pi\i\hbar T}
          \exp\bigg[-{m\over2\hbar\i T}(x''-x')^2\bigg]
  \tag\NUM.\num\\   \global\plus
  G^{(\bbbr)}(x'',x';E)&=\overh\sqrt{-{m\over2E}}
   \exp\bigg(-{\sqrt{-2mE}\over\hbar}\,\vert x''-x'\vert\bigg).
  \tag\NUM.\num\endalign$$
Insertion of $G^{(\bbbr)}(E)$ into Eq.~(\numbh) gives the Green
function of the motion in the half space $\bbbr^+$, which can be
transformed into the time-dependent propagator yielding
\plus
$$K^{(\bbbr^+)}(x'',x';T)
  =\sqrt{m\over2\pi\i\hbar T}\bigg[
  \exp\bigg(-{m\over2\i\hbar T}(x''-x')^2\bigg)-
  \exp\bigg(-{m\over2\i\hbar T}(x''+x')^2\bigg)\bigg]\enspace,
  \tag\NUM.\num$$
which is the classical result, e.g.\  [\FH, \SCHU].

\subsubsection{The Harmonic Oscillator}
In order to apply our formalism we have to know the Green
function of the harmonic oscillator (HO). It is not difficult to derive
form the propagator
\plus
$$\multline
  \int\limits_{x(t')=x'}^{x(t'')=x''}\CD x(t)
  \exp\Bigg[{\i m\over2\hbar}\int_{t'}^{t''}\Big(\dot x^2
       -\omega^2x^2\Big)dt\Bigg]
  \\
  =\sqrt{m\omega\over2\pi\i\hbar\sin\omega T}
  \exp\Bigg\{{\i m\omega\over2\hbar}\bigg[({x'}^2+{x''}^2)\cot\omega T
              -{2x'x''\over\sin\omega T}\bigg]\Bigg\}\enspace.
  \endmultline
  \tag\NUM.\num$$
the corresponding Green function $G^{(HO)}(E)$, which has the form
[\BAVE]
\plus
$$\multline
  G^{(HO)}(x'',x';E)=-\sqrt{m\over\pi\hbar^3\omega}\,
     \Gamma\bigg(\half-{E\over\hbar\omega}\bigg)
  \\   \times
  D_{-\half+{E\over\hbar\omega}}
  \bigg(\sqrt{2m\omega\over\hbar}\,x_>\bigg)
   D_{-\half+{E\over\hbar\omega}}
  \bigg(-\sqrt{2m\omega\over\hbar}\,x_<\bigg)\enspace.
  \endmultline
  \tag\NUM.\num$$
Here one can make use of the integral representation [\GRA, p.729],
$a_1>a_2$
\plus
$$\multline
  \int_0^\infty\coth^{2\nu}{x\over2}\exp\bigg[
                               -{a_1+a_2\over2}\cosh x\bigg]
  I_{2\mu}(\sqrt{a_1a_2}\sinh x)dx
  \\
  ={\Gamma(\half+\mu-\nu)\over\sqrt{a_1a_2}\,\Gamma(1+2\mu)}
  W_{\nu,\mu}(a_1)M_{\nu,\mu}(a_2)\enspace
  \endmultline
  \tag\NUM.\num$$
\edef\numdb{\NUM.\num}%
and exploit some relations between the Whittaker functions
$W_{\nu,\mu}(z)$ and $M_{\nu,\mu}(z)$, and the parabolic
cylinder-functions $D_\nu(z)$, respectively. The Green functions of
the harmonic oscillator in the half-spaces $x>a$ and $x<a$ are then
given by Eq.(\numbh) (compare also the case of the so-called double
oscillator [\BAVE]), respectively Eqs.~(\numbk,\numbl), and the bound
state energy levels are determined by
$$\alignat 3
  &D_{-\half+{E_n\over\hbar\omega}}
  \bigg(-\sqrt{2m\omega\over\hbar}\,a\bigg)=0
  &\qquad
  &\hbox{for $x<a$}\enspace,
  \tag\NUM.\num\\   \global\plus
  &D_{-\half+{E_n\over\hbar\omega}}
  \bigg(\sqrt{2m\omega\over\hbar}\,a\bigg)=0
  &\qquad
  &\hbox{for $x>a$}\enspace.
  \tag\NUM.\num\endalignat$$

\subsubsection{Morse Potential and Liouville Quantum Mechanics}
As the next example for motion in a half-space we consider the Morse
potential (M)
\plus
$$V^{(M)}(x)={V_0^2\hbar^2\over2m}(\e^{2x}-2\alpha\e^x)
  \tag\NUM.\num$$
($V_0>0$, $\alpha\in\bbbr$ constants). This potential has
$N_{Max}<\alpha V_0-\half$ bound states. The Green function can be
explicitly calculated by the path integral formalism
[\CIWI, \DURa, \GROb, \PAKSa] and has the form
\plus
$$\myalign
  &\ih                    \int_0^\infty dT\,\e^{\i ET/\hbar}
  \int\limits_{x(t')=x'}^{x(t'')=x''}\CD x(t)
  \exp\left\{\ih\int_{t'}^{t''}\bigg[{m\over2}\dot x^2
  -{V_0^2\hbar^2\over2m}(\e^{2x}-2\alpha\e^x)\bigg]dt\right\}
  \\   &
  ={m\over V_0\hbar^2}
  {\Gamma(\half+\sqrt{-2mE}/\hbar-\alpha V_0)\over
   \Gamma(1+\sqrt{-8mE}/\hbar)\e^{(x'+x'')/2}}
  W_{\alpha V_0,{\sqrt{-2mE}\over\hbar}}\big(2V_0 \e^{x_>}\big)
  M_{\alpha V_0,{\sqrt{-2mE}\over\hbar}}\big(2V_0 \e^{x_<}\big)
  \enspace.
  \tag\NUM.\num\endalign$$
The Green functions in the half-spaces $x>a$ and $x<a$ are then given
by Eq.~(\numbh), respectively Eqs.~(\numbk,\numbl), and the bound state
energy levels are determined by
$$\alignat 3
  &M_{\alpha V_0,\sqrt{-2mE_n}/\hbar}(2V_0 \e^a)=0
  &\qquad
  &\hbox{for $x<a$}\enspace,
  \tag\NUM.\num\\   \global\plus
  &W_{\alpha V_0,\sqrt{-2mE_n}/\hbar}(2V_0 \e^a)=0
  &\qquad
  &\hbox{for $x>a$}\enspace.
  \tag\NUM.\num\endalignat$$
Liouville (L) quantum mechanics can be studied by setting
$\alpha=0$ in $V^{(M)}$. We have [\GRSa]
\plus
$$G^{(L)}(x'',x';E)={2m\over\hbar^2}
  I_{\sqrt{-2mE}/\hbar}(V_0 \e^{x_<})
  K_{\sqrt{-2mE}/\hbar}(V_0 \e^{x_>})\enspace.
  \tag\NUM.\num$$
Again, the bound state energy-levels $E_n$ in the half-space $x>a$ are
determined by
\plus
$$K_{\sqrt{-2mE_n}/\hbar}(V_0 \e^a)=0\enspace.
  \tag\NUM.\num$$
Note that the function $K_{\i\nu}(z)$ for large values of $\nu$ is an
oscillating $\sin(z)$-like function (e.g.\ [\MOS], p.142).

\subsubsection{The Wood Saxon Potential}
We consider the Wood-Saxon potential. It is defined as [\FLU]
($b,R,V_0>0$ constants)
\plus
$$V(r)=-{V_0\over 1+\e^{(r-b)/R}},\qquad r>0\enspace.
  \tag\NUM.\num$$
The corresponding radial path integral formulation has the form
($D=3$)
\plus
$$K_l(r'',r';T)=\int\limits_{r(t')=r'}^{r(t'')=r''}
  \mu_{l+\half}[r^2]\CD r(t)\exp\left\{\ih\int_{t'}^{t''}
  \bigg[{m\over2}\dot r^2+{V_0\over 1+\e^{(r-b)/R}}\right]dt\bigg\}
  \enspace.
  \tag\NUM.\num$$
This potential is used to describe the potential trough for the strong
interaction near the nucleus for radial symmetric nuclei. There are
only a finite number of bound states. Putting $l=0$ makes this path
integral looking like the ``smooth step'' potential in $\bbbr$ with a
barrier at $r=0$. The path integral solution of the ``smooth step''
potential in the entire $\bbbr$ in turn is given by [\GROe, \KLEMUS]
\plus
$$\myalign
   \ih&                   \int_0^\infty dT\,\e^{\i ET/\hbar}
  \int\limits_{x(t')=x'}^{x(t'')=x''}\CD x(t)
  \exp\left\{\ih\int_{t'}^{t''}\bigg[{m\over2}\dot x^2
      +{V_0\over 1+\e^{(x-b)/R}}\bigg]dt\right\}
  \\   &
  ={2mR\over\hbar^2}{\Gamma(m_1)\Gamma(m_1+1)\over
           \Gamma(m_1+m_2+1)\Gamma(m_1-m_2+1)}
  \\   &\qquad\times
  \bigg({1-\tanh{x_<-b\over2R}\over2}\bigg)^{m_1-m_2\over2}
  \bigg({1+\tanh{x_<-b\over2R}\over2}\bigg)^{m_1+m_2\over2}
  \\   &\qquad\times
  \bigg({1-\tanh{x_>-b\over2R}\over2}\bigg)^{m_1-m_2\over2}
  \bigg({1+\tanh{x_>-b\over2R}\over2}\bigg)^{m_1+m_2\over2}
  \\   &\qquad\times
  {_2}F_1\bigg(m_1,m_1+1;m_1-m_2+1;{1-\tanh{x_>-b\over2R}\over2}\bigg)
  \\   &\qquad\times
  {_2}F_1\bigg(m_1,m_1+1;m_1+m_2+1;{1+\tanh{x_<-b\over2R}\over2}\bigg)
  \enspace.
  \tag\NUM.\num\endalign$$
Here denote $m_{1,2}=\sqrt{2m}\,R\big(\sqrt{-E-V_0}\pm\sqrt{-E}\,\big)/
\hbar$. With a barrier at $x=a$, such that we consider motion in the
half-space $x>a$, we obtain that the Green function of the Wood-Saxon
potential is given by Eq.~(\numbh), and the bound state energy levels
are determined by (with $0<\vert E_n\vert<V_0$)
\plus
$${_2}F_1\bigg(\beta+\i\lambda,\beta+\i\lambda+1;1+2\beta;
  {1-\tanh{a-b\over2R}\over2}\bigg)=0\enspace.
  \tag\NUM.\num$$
Here denote $\beta^2=-2mE_nR^2/\hbar^2$, $\lambda^2=2m(E_n+V_0)R^2/
\hbar^2$. The claim of Duru [\DURe] of having solved this problem cannot
be seen as correct due to an improper handling of the boundary-conditions.

\subsubsection{The Linear Potential}
The problem of the linear potential (L) $V(x)=kx$ in the half-space
$x>0$ has a well-known result, i.e.\ the infinite number of bound state
energy levels are determined by the zeros of the Airy-function [\FLU].
The propagator of the linear potential in the entire $\bbbr$ is quite
easy to obtain due to the fact that it belongs to the class of quadratic
Lagrangians whose path integral solution are determined by the
classical action. It is given by [\FH, \SCHU] ($x\in\bbbr$)
\plus
$$\multline
  \int\limits_{x(t')=x'}^{x(t'')=x''}\CD x(t)
  \exp\left[\ih\int_{t'}^{t''}
  \bigg({m\over2}\dot x^2-kx\bigg)dt\right]
  \\   =
  \bigg({m\over2\pi\i\hbar T}\bigg)^{1/2}
  \exp\bigg[\ih\bigg({m\over2}{(x''-x')^2\over T}
      -{kT\over2}(x'+x'')-{k^2T^3\over24m}\bigg)\bigg].
  \endmultline
  \tag\NUM.\num$$
The corresponding energy dependent Green function cannot be evaluated
in an obvious direct way by means of a Fourier transformation
from the propagator due to the rather nasty $T^3$-dependence. However,
as it is shown in Appendix B, a space-time transformation does the job.
For the Green function we obtain
\plus
$$\multline
  G^{(L)}(x'',x';E)={4\over3}{m\over\hbar^2}\bigg[
   \bigg(x'-{E\over k}\bigg)\bigg(x''-{E\over k}\bigg)\bigg]^{1/2}
   \\   \times
   K_{1/3}\left[{2\over3}\sqrt{2mk\over\hbar^2}
   \bigg(x_>-{E\over k}\bigg)^{3/2}\right]
   I_{1/3}\left[{2\over3}\sqrt{2mk\over\hbar^2}
   \bigg(x_<-{E\over k}\bigg)^{3/2}\right]\enspace.
  \endmultline
  \tag\NUM.\num$$
The Green functions in the half-spaces $x>a$ and $x<a$ are then given
by Eq.~(\numbh), respectively Eqs.~(\numbk,\numbl), and the bound state
energy levels are determined by $(n\in\bbbn_0$, $x>a$)
\plus
$$K_{1/3}\left[{2\over3}\sqrt{2mk\over\hbar^2}
   \bigg(a-{E_n\over k}\bigg)^{3/2}\right]
  =\Ai\left[\bigg(a-{E_n\over k}\bigg)
       \bigg({2mk\over\hbar^2}\bigg)^{1/3}\right]=0\enspace,
  \tag\NUM.\num$$

\subsection{Boundary Problems in Boxes}
\subsubsection{The Infinite Potential Well}
Again we start with a simple example, i.e. with the infinite potential
well (IPW). We need the Green function of the free particle, and insert
it into Eq.(\numbf). Performing the Fourier transformation with respect
to $E$ then gives the well-known result [\SCHU] ($-b<x<b$)
\plus
$$\myalign
  K^{(IPW)}(x'',x';T)&=\sqrt{m\over2\pi\i\hbar T}
  \\    \times
  \sum_{n=-\infty}^\infty
  \Bigg\{\exp\bigg[-&{m\over2\i\hbar T}(x''-x'+4nb)^2\bigg]
     -\exp\bigg[
  -{m\over2\i\hbar T}(x''+x'+2(2n+1)b)^2\bigg]\Bigg\}
  \\   &={1\over4b}\Bigg[\Theta_3
  \bigg({x''-x'\over4b},-{\pi\hbar T\over8mb^2}\bigg)
   -\Theta_3\bigg({x''+x'\over4b}+\half,-{\pi\hbar T\over8mb^2}\bigg)
  \Bigg]
  \\   &={1\over b}\sum_{n=1}^\infty
    \exp\bigg(-\i\hbar T{\pi^2n^2\over8mb^2}\bigg)
    \cos\bigg({\pi n\over2b}x'\bigg)\cos\bigg({\pi n\over2b}x''\bigg).
  \tag\NUM.\num\endalign$$
Here $\Theta_3$ denotes a Jacobi $\Theta$-function.

\subsubsection{The Harmonic Oscillator}
This particular problem has attracted some attention by
Aguilera-Navarro et al.\ [\ANIKZ], Consortini and Frieden [\COFRI],
Marin and Cruz [\MACR], and Vawter [\VAW]. By means of the Green
function for the harmonic oscillator we obtain for the Green function
in the box $a<x<b$ Eq.~(\numbi), and the bound state energy-levels
are determined by
\plus
$$\multline
  D_{-\half+{E_n\over\hbar\omega}}
  \bigg(\sqrt{2m\omega\over\hbar}\,a\bigg)
  D_{-\half+{E_n\over\hbar\omega}}
  \bigg(-\sqrt{2m\omega\over\hbar}\,b\bigg)
  \\
  =D_{-\half+{E_n\over\hbar\omega}}
  \bigg(\sqrt{2m\omega\over\hbar}\,b\bigg)
  D_{-\half+{E_n\over\hbar\omega}}
  \bigg(-\sqrt{2m\omega\over\hbar}\,a\bigg)\enspace.
  \endmultline
  \tag\NUM.\num$$

\subsection{Boundary Problems in Radial Boxes}
\subsubsection{The Radial Box}
In order to treat the problem of the simple radial box we consider
the radial free Green function which is given by
\plus
$$\myalign
     &(r'r'')^{3-D\over2}\ih\int_0^\infty dT\,\e^{\i ET/\hbar}
  \int\limits_{r(t')=r'}^{r(t'')=r''}\CD r(t)
  \exp\left[\ih\int_{t'}^{t''}
  \bigg({m\over2}\dot r^2-\hbar^2{l(l+D-2)\over2mr^2}\bigg)dt\right]
  \\   &
  =(r'r'')^{3-D\over2}\ih\int_0^\infty dT\,\e^{\i ET/\hbar}
  \int\limits_{r(t')=r'}^{r(t'')=r''}\mu_{l+{D-2\over2}}[r^2]\CD r(t)
  \exp\left({\i m\over2\hbar}\int_{t'}^{t''}\dot r^2dt\right)
  \\   &
  ={2m\over\hbar^2(r'r'')^{2-D\over2}}
  K_{l+{D-2\over2}}\bigg(\sqrt{-2mE}\,{r_>\over\hbar}\bigg)
  I_{l+{D-2\over2}}\bigg(\sqrt{-2mE}\,{r_<\over\hbar}\bigg)\enspace.
  \tag\NUM.\num\endalign$$
\edef\numdc{\NUM.\num}%
Therefore the Green function for the simple radial box $0<r<R$ is
given by Eq.~(\numbi) and the energy levels are determined by
\plus
$$I_{l+{D-2\over2}}\bigg(\sqrt{-2mE_n}\,{R\over\hbar}\bigg)=0\enspace,
  \tag\NUM.\num$$
i.e.\ by the zeros $z_{\lambda,n}$ of the modified Bessel function
$I_\lambda(z)$. The corresponding problem with $r>R$ describes the free
motion in $D$  dimensions with a impenetrable (hard) sphere at the
origin. There are no bound states.

This result allows us to describe the problem of the radial potential
\plus
$$V_\lambda(r)={\hbar\over2m}{\lambda^2-\viert\over2mr^2}
  \tag\NUM.\num$$
within, respectively outside, a radial box. The Green function of the
problem without boundary is given by [it can be considered as an analytic
continuation of Eq~(\numdc)]:
\plus
$$G^{(V_\lambda)}(r'',r';E)=
  {2m\over\hbar^2}
  K_{\lambda}\bigg(\sqrt{-2mE}\,{r_>\over\hbar}\bigg)
  I_{\lambda}\bigg(\sqrt{-2mE}\,{r_<\over\hbar}\bigg)\enspace,
  \tag\NUM.\num$$
and the Green function with boundary in the space $0<r<R$ is given by
Eq.~(\numbi), and the bound state energy levels are determined by
\plus
$$I_{\lambda}\bigg(\sqrt{-2mE_n}\,{R\over\hbar}\bigg)=0\enspace.
  \tag\NUM.\num$$
This kind of problems appear e.g.\ by considering a sector of angel
$\alpha$ which has a boundary at $r=R$, where $\lambda=l\pi/\alpha$
($l\in\bbbn$), i.e.~we have motion inside a region which looks like a
piece of a cake [\CCH].

Vice versa, we obtain for the Green function for the potential
$V_\lambda$ in the space $r>R$
\plus
$$\multline
  G^{(r>R)}(r'',r';E)={2m\over\hbar^2}
  K_{\lambda}\bigg(\sqrt{-2mE}\,{r_>\over\hbar}\bigg)
  I_{\lambda}\bigg(\sqrt{-2mE}\,{r_<\over\hbar}\bigg)
  \\
  -{2m\over\hbar^2}
  {I_{\lambda}\bigg(\sqrt{-2mE}\,\dfrac{R}\hbar\bigg)\over
   K_{\lambda}\bigg(\sqrt{-2mE}\,\dfrac{R}\hbar\bigg)}
   K_{\lambda}\bigg(\sqrt{-2mE}\,{r'\over\hbar}\bigg)
   K_{\lambda}\bigg(\sqrt{-2mE}\,{r''\over\hbar}\bigg)\enspace.
  \endmultline
  \tag\NUM.\num$$
Hence, we describe the problem of the potential $V_\lambda$ outside a
hard sphere (hard disc) of radius $R$. A model for such a Green
function is the Green function for a Bohm-Aharonov solenoid outside a
hard disc of radius $R$ with flux $\Phi$ inside. Then $\lambda=\vert\nu
+q\Phi/2\pi c\hbar\vert$ ($\nu\in\bbbz$), with $q$ the charge of the
particle and $c$ the velocity of light (e.g.\ [\INO, \INOSa, \WIE] and
references therein).

\subsubsection{The Radial Harmonic Oscillator}
The model of the radial harmonic oscillator (RHO) can either be seen as
a radial harmonic oscillator inside a radial box, or a radial harmonic
oscillator with a hard sphere at the origin. We have for the path
integral solution for the radial harmonic oscillator [\DURd, \GOO, \PI]
\plus
$$\myalign
  &\int\limits_{r(t')=r'}^{r(t'')=r''}\CD r(t)
  \exp\left[\ih\int_{t'}^{t''}\bigg({m\over2}\dot r^2
  -\hbar^2{l(l+D-2)\over2mr^2}
           -{m\over2}\omega^2r^2\bigg)dt\right]
  \\   &
  =\int\limits_{r(t')=r'}^{r(t'')=r''}
  \mu_{l+{D-2\over2}}[r^2]\CD r(t)
  \exp\left[{\i m\over2\hbar}\int_{t'}^{t''}
  \big(\dot r^2-\omega^2r^2\Big)dt\right]
  \\   &
  =\sqrt{r'r''}{m\omega\over \i\hbar\sin\omega T}
  \exp\bigg[-{m\omega\over2\i\hbar}({r'}^2+{r''}^2)\cot\omega T\bigg]
  I_{l+{D-2\over2}}
  \bigg({m\omega r'r''\over \i\hbar\sin\omega T}\bigg)\enspace.
  \tag\NUM.\num\endalign$$
\edef\numda{\NUM.\num}%
The corresponding Green function is given by [\BAVE, \BVK]
(c.f.\ Eq.~(\numdb))
\plus
$$\multline
  G^{(RHO)}_l(r'',r';E)={\Gamma[\half(l+{D\over2}
   -{E\over\hbar\omega})]\over\hbar\omega(r'r'')^{D/2}\Gamma(l+D/2)}
  \\   \times
  W_{{E\over2\hbar\omega},\half(l+{D-2\over2})}
   \bigg({m\omega\over\hbar}r_>^2\bigg)
  M_{{E\over2\hbar\omega},\half(l+{D-2\over2})}
   \bigg({m\omega\over\hbar}r_<^2\bigg)
  \enspace.
  \endmultline
  \tag\NUM.\num$$
With a barrier at $x=R$ we obtain the Green function of the radial
harmonic oscillator for a radial box by Eq.~(\numbi), and the bound
state energy levels are determined by
$$\alignat 3
  &M_{{E_n\over2\hbar\omega},\half(l+{D-2\over2})}
   \bigg({m\omega\over\hbar}R^2\bigg)=0
  &\qquad
  &\hbox{for $r<R$}\enspace,
  \tag\NUM.\num\\   \global\plus
  &W_{{E_n\over2\hbar\omega},\half(l+{D-2\over2})}
   \bigg({m\omega\over\hbar}R^2\bigg)=0
  &\qquad
  &\hbox{for $r>R$}\enspace.
  \tag\NUM.\num\endalignat$$

\subsubsection{The Coulomb Potential}
We consider the Coulomb potential (C). The path integral reads
($\vec x\in\bbbr^D$)
\plus
$$K^{(C)}(\vec x'',\vec x';T)=
  \int\limits_{\vec x(t')=\vec x'}^{\vec x(t'')=\vec x''}\CD \vec x(t)
  \exp\left\{\ih\int_{t'}^{t''}
  \bigg[{m\over2}\dot{\vec x^2}+{q_1q_2\over\vert\vec x\vert}
  \bigg]dt\right\}\enspace.
  \tag\NUM.\num$$
Here only the Green function  can be explicitly calculated. The
radial Coulomb Green function has the form (see e.g.\ [\BAVE, \CHc,
\GROm, \GRSf, \GRSg, \STEc] and references therein)
\plus
$$\multline
  G^{(C)}_l(r'',r';E)=(r'r'')^{1-D\over2}\overh\sqrt{-{m\over2E}}\,
   {\Gamma\big(l+{D-1\over2}+\i p\big)\over(2l+D-2)! }
   \\   \times
   W_{-\i p,l+{D-2\over2}}\bigg(\sqrt{-{8mE\over\hbar^2}}\,r_>\bigg)
   M_{-\i p,l+{D-2\over2}}\bigg(\sqrt{-{8mE\over\hbar^2}}\,r_<\bigg)
  \enspace,
  \endmultline
  \tag\NUM.\num$$
with $p=(q_1q_2/\hbar)\sqrt{m/2E}$. With a barrier at $x=R$ we obtain
the Green function of the Coulomb potential for a radial box by
Eq.~(\numbi), and the bound state energy levels are determined by
$$\alignat 3
  &M_{{q_1q_2\over\hbar}\sqrt{-{m\over2E_n}},l+{D-2\over2}}
     \bigg(\sqrt{-8mE_n}\,{R\over\hbar}\bigg)=0
  &\qquad
  &\hbox{for $r<R$}\enspace,
  \tag\NUM.\num\\   \global\plus
  &W_{{q_1q_2\over\hbar}\sqrt{-{m\over2E_n}},l+{D-2\over2}}
     \bigg(\sqrt{-8mE_n}\,{R\over\hbar}\bigg)=0
  &\qquad
  &\hbox{for $r>R$}\enspace.
  \tag\NUM.\num\endalignat$$

\subsection{Boundary Problems in Radial Rings}
It is not difficult to generalize the preceeding results to the case
of radial rings. The corresponding Green function in a radial ring
$a<r<R$ are given by Eq.~(\numbj) and the bound state energy levels are
determined by:

\subsubsection{Simple Radial Ring}
\plus
$$\multline
  I_{l+{D-2\over2}}\bigg(\sqrt{-2mE_n}\,{a\over\hbar}\bigg)
  K_{l+{D-2\over2}}\bigg(\sqrt{-2mE_n}\,{R\over\hbar}\bigg)
  \\  =
  I_{l+{D-2\over2}}\bigg(\sqrt{-2mE_n}\,{R\over\hbar}\bigg)
  K_{l+{D-2\over2}}\bigg(\sqrt{-2mE_n}\,{a\over\hbar}\bigg)
  \enspace.
  \endmultline
  \tag\NUM.\num$$

\subsubsection{Radial Harmonic Oscillator in a Ring}
\plus
$$\multline
  M_{{E_n\over2\hbar\omega},\half(l+{D-2\over2})}
   \bigg({m\omega\over\hbar}a^2\bigg)
  W_{{E_n\over2\hbar\omega},\half(l+{D-2\over2})}
   \bigg({m\omega\over\hbar}R^2\bigg)
  \\  =
  W_{{E_n\over2\hbar\omega},\half(l+{D-2\over2})}
   \bigg({m\omega\over\hbar}a^2\bigg)
  M_{{E_n\over2\hbar\omega},\half(l+{D-2\over2})}
   \bigg({m\omega\over\hbar}R^2\bigg)
  \enspace.
  \endmultline
  \tag\NUM.\num$$

\subsubsection{Coulomb Potential in a Ring}
\plus
$$\multline
  M_{{q_1q_2\over\hbar}\sqrt{-{m\over2E_n}},l+{D-2\over2}}
     \bigg(\sqrt{-8mE_n}\,{a\over\hbar}\bigg)
  W_{{q_1q_2\over\hbar}\sqrt{-{m\over2E_n}},l+{D-2\over2}}
     \bigg(\sqrt{-8mE_n}\,{R\over\hbar}\bigg)
  \\  =
  W_{{q_1q_2\over\hbar}\sqrt{-{m\over2E_n}},l+{D-2\over2}}
     \bigg(\sqrt{-8mE_n}\,{a\over\hbar}\bigg)
  M_{{q_1q_2\over\hbar}\sqrt{-{m\over2E_n}},l+{D-2\over2}}
     \bigg(\sqrt{-8mE_n}\,{R\over\hbar}\bigg)
  \enspace.
  \endmultline
  \tag\NUM.\num$$

\subsection{Moving Boundary Problems}
\subsubsection{The Infinite Potential Well}
We consider the example of the infinite potential well with one boundary
fixed at $x=0$, and the other moving uniformly in time according to
$L(t)=L_0 \zeta(t)$ [\DMN]. According to Eq.~(\numca), the result then
has for $\omega'=0$ the form
\plus
$$\multline
  K^{(IPW)}(x'',x';t'',t')
  ={(\zeta'\zeta'')^{-1/2}\over2L_0}
  \exp\left[{\i m\over2\hbar}\left({x''}^2{\dot\zeta''\over\zeta''}
                     -{x'}^2{\dot\zeta'\over\zeta'}\right)\right]
  \\    \times
  \Bigg[\Theta_3\bigg({x''/\zeta''-x'/\zeta'\over2L_0}
          ,-{\pi\hbar\tau(t'')\over2mL_0^2}\bigg)
   -\Theta_3\bigg({x''/\zeta''+x'/\zeta'\over2L_0}
          ,-{\pi\hbar\tau(t'')\over2mL_0^2}\bigg)\Bigg]\enspace.
  \endmultline
  \tag\NUM.\num$$

\subsubsection{Moving Boundary in the Half-Space}
We consider the half-space with the boundary moving uniformly in time
according to Eq.~(\numcb) and we consider $f(t)=vt$.
Hence $\ddot f=0$, and for a free particle we obtain
\plus
$$\multline
  K^{(MB)}(x'',x';t'',t')
  =\sqrt{m\over2\pi\i\hbar T}\,\exp\left\{{\i m\over\hbar}
  \bigg[v(x''-vt'')-v(x'-vt')+{v^2\over2}T\bigg]\right\}
  \\   \times
  \Bigg[\exp\bigg({\i m\over2\hbar T}\vert x''-x'-vT\vert^2\bigg)
  -\exp\bigg({\i m\over2\hbar T}\vert x''+x'-v(t''+t')\vert^2\bigg)
  \Bigg]\enspace.
  \endmultline
  \tag\NUM.\num$$
Generally, if the ``mirror'' principle can be applied, say for the
harmonic oscillator, the result reads
\plus
$$\multline
  K^{(MB)}(x'',x';t'',t')
  =\exp\Bigg\{{\i m\over\hbar}
  \bigg[v(x''-vt'')-v(x'-vt')+{v^2\over2}T\bigg]\Bigg\}
  \\  \times
  \Big[K_0(x''-vt'',x'-vt';T)-K_0(x''-vt'',-(x'-vt');T)\Big]
  \enspace.
  \endmultline
  \tag\NUM.\num$$

%----------------------------------------------------------------------
%                          END OF FILE4
%----------------------------------------------------------------------

\PLUS
\glno=0                      %V
\section{Summary}
In this paper I have presented a comprehensive approach to the theory of
boundary problems with Dirichlet boundary-conditions derived by an
appropriate limiting process from $\delta$-function perturbations in
path integrals.

First, I could derive from a path integral formulation and an exact
summation of a perturbation expansion, closed formul\ae\ for the
problem of multiple $\delta$-function perturbations in one dimension.
This result could be used to derive an integral representation for the
problem of multiple perpendicular $\delta$-function perturbations along
lines, planes and hyperplanes in $D$ dimensions, in the form of
$(D-1)$-fold convolutions of the corresponding one-dimensional problem.

I could also discuss the problems of a $\delta$-function perturbation
along a line, respectively its equivalent of two particles interacting
via a $\delta$-function, and a moving $\delta$-function perturbation.

Second, I could consider boundary problems with Dirichlet
boundary-conditions. They were derived form the ones with
$\delta$-function perturbations by letting the strength of the
$\delta$-function perturbations be infinite repulsive. This gave closed
expressions for the one-dimensional motions in half-spaces, boxes, radial
boxes, rings and moving boundaries. The advanced problem of multidimensional
perpendicular boundaries along perpendicular lines and planes was given
similarly as for the $\delta$-function perturbations, i.e.\ by
convolutions of the corresponding Green functions of the one-dimensional
case, however, only two boundaries are required for
each dimension, thus simplifying the corresponding one-dimensional
Green functions, and the number of boundaries must not exceed
$(2\times\hbox{dimension})$.

These models of multiple $\delta$-function perturbations and boundaries,
respectively, along perpendicular lines and planes should serve as a
numerically simple approach for the investigation of band structures.

Several examples put our results in a better context and illustrated
the formalism. Here some long unsolved problems in path integration
could be treated successfully, for instance the Green function of the linear
potential, and the Wood-Saxon potential in the half-space $x>a$.
For the former, there are an infinite number of levels determined by
the zeros of the Airy-function, whereas for the latter there are only
a finite number of levels. For the linear potential it was required to
calculate the Green function in the entire $\bbbr$ which was done
in Appendix B.

However, the question arises how to deal with other than Dirichlet
boundary-conditions [\CAR, \CFG, \CMS], which present only one
possibility to define a self-adjoint extension of the (free) Hamiltonian
with boundaries. It is known that an
appropriate chosen $\delta$-function perturbation can do the job; at
least in case of the one-dimensional free particle in $\bbbr^+$ the
propagator with general boundary-conditions at $x=0$ can be explicitly
computed [\CMS]. It is derived from two perturbed constructively
interfering kernels, i.e.\ a modified ``mirror'' principle is applied.
As we know, this simple method does not work in the general case. All what we
know in the general case is the corresponding Green function for the
perturbed-, respectively boundary-problem. Of course, the same
difficulties arise for the case of more than one boundary and in
higher dimensions. These problems, however, will be discussed elsewhere.

\vfill\noindent
{\bf Acknowledgements}
\par\noindent
I would like to thank the members of the II.~Institut f\"ur Theoretische
Physik, Hamburg University, where part of this work was done, for their
kind hospitality.

%----------------------------------------------------------------------
%                          END OF FILE4
%----------------------------------------------------------------------

\eject\noindent
\Chapno=1
\glno=0                      %A
\appendix{Green Function for Multiple $\delta$-Function Perturbations}
In this appendix I want to show Eq.~(\numbd).
The proof goes by induction. It is obvious that it is sufficient to
show the validity of Eq.~(\numbd) for the denominator.
Let us consider now the determinant ($n\in\bbbn$)
$\det(a_{ij})=0$, with the $(n-1)\times(n-1)$ matrix
\plus
$$(a_{ij})
   =\bigg(\GV(a_i,a_j;E)-{\delta_{ij}\over\gamma_i}\bigg)\enspace.
  \tag\AA.\num$$
We know that Eq.~(\numbd) is valid for $n=1,2$.
We make the step form $n-1$ to $n$. Doing this, every matrix entry
is changed in the following way
\plus
$$a_{ij}\mapsto\widetilde{a_{ij}}=
  {a_{nn}a_{ij}-a_{in}a_{nj}\over a_{nn}}\enspace.
  \tag\AA.\num$$
Therefore we obtain by expanding the determinant
\plus
$$\myalign
  \!\!\!\!\!\!
  &\det\Big(\widetilde{a_{ij}}\big\vert_{i,j=1}^{n-1}\Big)
  \\   &
  ={1\over a_{nn}^{n-1}}
  \left\vert\matrix
  a_{11}a_{nn}-a_{1n}a_{n1}
         &\hdots  &a_{1,n-1}a_{nn}-a_{1n}a_{n,n-1} \\
  \vdots &\ddots  &\vdots                          \\
  a_{n-1,1}a_{nn}-a_{n-1,n}a_{n1}
         &\hdots  &a_{n-1,,n-1}a_{nn}-a_{n-1,n}a_{n,n-1}
  \endmatrix\right\vert
  \\   &
  ={1\over a_{nn}^{n-1}}\left\{
   a_{nn}^{n-1}
  \left\vert\matrix
  a_{11}    &a_{12}    &\hdots &a_{1,n-1}   \\
  a_{21}    &a_{22}    &\hdots &a_{2,n-1}   \\
  \vdots    &\vdots    &\ddots &\vdots      \\
  a_{n-1,1} &a_{n-1,2} &\hdots &a_{n-1,n-1} \endmatrix\right\vert
  \right.
  \\   &\quad
  -a_{nn}^{n-2}a_{n,n-1}
  \left\vert\matrix
  a_{11}    &\hdots &a_{1,n-2}   &a_{1,n}   \\
  a_{21}    &\hdots &a_{2,n-2}   &a_{2,n}   \\
  \vdots    &\ddots &\vdots      &\vdots    \\
  a_{n-1,1} &\hdots &a_{n-1,n-2} &a_{n-1,n} \endmatrix\right\vert
  \\   &\quad+\hdots
  +(-1)^{n}a_{nn}^{n-2}a_{n,1}
  \left\vert\matrix
  a_{1n}    &a_{1,2}   &\hdots &a_{1,n-1}   \\
  a_{2n}    &a_{2,2}   &\hdots &a_{2,n-1}   \\
  \vdots    &\vdots    &\ddots &\vdots      \\
  a_{n-1,n} &a_{n-1,2} &\hdots &a_{n-1,n-1} \endmatrix\right\vert
  \\   &\quad
  +a_{nn}^{n-3}a_{n,n-1}a_{n,n-2}
  \left\vert\matrix
  a_{11}    &\hdots &a_{1,n}   &a_{1,n}   \\
  a_{21}    &\hdots &a_{2,n}   &a_{2,n}   \\
  \vdots    &\ddots &\vdots    &\vdots    \\
  a_{n-1,1} &\hdots &a_{n-1,n} &a_{n-1,n} \endmatrix\right\vert
  \\   &\quad+\hdots
  +(-1)^{n-1}(a_{n,n-1}\hdots a_{n,1})
  \left.\left\vert\matrix
  a_{1n}    &\hdots &a_{1j}    &\hdots  &a_{1,n}     \\
  a_{2n}    &\hdots &a_{2j}    &\hdots  &a_{2,n}     \\
  \vdots    &\ddots &\vdots    &\ddots  &\vdots      \\
  a_{n-1,n} &\hdots &a_{n-1,j} &\hdots  &a_{n-1,n}
  \endmatrix\right\vert\enspace\right\}\enspace.
  \tag\AA.\num\endalign$$
The last terms with the power of the $a_{nn}$ factor less than
$n-2$ are all zero because two columns  are always equal.
The last non-vanishing term can be rearranged by a permutation where the
first column is put into the last with increasing second index of the
$a_{1,j}$, etc.
The emerging sum is now the definition of the expansion of a
$n\times n$ matrix with the required feature, i.e.\ we have
\plus
$$\det\Big(\widetilde{a_{ij}}\big\vert_{i,j=1}^{n-1}\Big)
  =\det\Big(a_{ij}\big\vert_{i,j=1}^n\Big)\enspace,
  \tag\AA.\num$$
which shows the induction.

%----------------------------------------------------------------------
%                          END OF FILEA
%----------------------------------------------------------------------

\PLUS
\glno=0                      %B
\appendix{Green Function for the Linear Potential}
We consider the path integral formulation for the linear potential
\plus
$$K(x'',x';T)=
  \int\limits_{x(t')=x'}^{x(t'')=x''}\CD x(t)\exp\left[\ih
  \int_{t'}^{t''}\bigg({m\over2}\dot x^2-kx\bigg)dt\right]\enspace.
  \tag\BB.\num$$
We perform the combined space-time transformation (c.f.\ Refs.~[\GRSb,
\KLE, \STEa] and references therein)
\plus
$$y=\bigg(x-{E\over k}\bigg)^{3/2}\enspace,\qquad
  dt={4\over9}y^{-2/3}ds\enspace,
  \tag\BB.\num$$
and introduce a new pseudo-time $s''$
\plus
$$s''={9\over4}\int_{t'}^{t''}y^{2/3}(t)dt\enspace,
  \tag\BB.\num$$
in this path integral, where the lattice implementation is given by
$\Delta t_{j}={4\over9}y^{-1/3}_{j}y^{-1/3}_{j-1}$
to guarantee a symmetric transformation with respect to initial and
final coordinates. This gives the transformation formul\ae
$$\myalign
  K(x'',x';T)&=\int_{-\infty}^\infty{dE\over2\pi\i}
   \e^{-\i ET/\hbar}G(x'',x';E)
  \tag\BB.\num\\    \global\plus
  G(x'',x';E)&=\ih{2\over3}(y'y'')^{-1/6}
  \int_0^\infty ds''\,\e^{-4\i s''k/9\hbar}
  \widetilde K(y'',y';s'')\enspace,
  \tag\BB.\num\endalign$$
with the transformed path integral $\widetilde K(s'')$ given by
\plus
$$\myalign
  \widetilde K(y'',y';s'')
  &=\int\limits_{y(0)=y'}^{y(s'')=y''}\CD y(s)\exp\Bigg[\ih
   \int_0^{s''}\bigg({m\over2}\dot y^2+{\hbar^2\over8my^2}{5\over9}
   \bigg)ds\Bigg]
  \\   &
  =\int\limits_{y(0)=y'}^{y(s'')=y''}\mu_{1/3}[y^2] \CD
   y(s)\exp\Bigg({\i m\over2\hbar}\int_0^{s''}\dot y^2ds\Bigg)
  \\   &
  =\sqrt{y'y''}{m\over\i\hbar s''}
   \exp\bigg[-{m\over2\i\hbar s''}({y'}^2+{y''}^2)\bigg]
   I_{1/3}\bigg({my'y''\over\i\hbar s''}\bigg)\enspace.
  \tag\BB.\num\endalign$$
Here use has been made of the radial path integral formulation
(\numbm) with the functional weight $\mu_\nu$ together with
the path integral solution (\numda) for inverse square radial
potentials. Making use of the integral representation [\GRA, p.719]
\plus
$$\int_0^\infty \e^{-a/x-bx}J_\nu(cx)dx
  =2J_\nu\bigg[\sqrt{2a\big(\sqrt{b^2+c^2}-b\big)}\,\bigg]
    K_\nu\bigg[\sqrt{2a\big(\sqrt{b^2+c^2}+b\big)}\,\bigg]
  \tag\BB.\num$$
we obtain for $G(E)$
\plus
$$\multline
  G(x'',x';E)={4\over3}{m\over\hbar^2}\bigg[
   \bigg(x'-{E\over k}\bigg)\bigg(x''-{E\over k}\bigg)\bigg]^{1/2}
   \\   \times
   K_{1/3}\left[{2\over3}\sqrt{2mk\over\hbar^2}
   \bigg(x_>-{E\over k}\bigg)^{3/2}\right]
   I_{1/3}\left[{2\over3}\sqrt{2mk\over\hbar^2}
   \bigg(x_<-{E\over k}\bigg)^{3/2}\right]\enspace,
  \endmultline
  \tag\BB.\num$$
(compare Steiner [\STEa] for the case $E=0$). The modified Bessel
functions $I_{1/3}$ and $K_{1/3}$ can be rewritten as Airy functions.
One has to make use of the following relations [\MOS, p.75]
$$\myalign
  \Ai(z)&=\bigg({z\over9}\bigg)^{1/2}
  \Bigg[I_{-1/3}\bigg({2\over3}z^{3/2}\bigg)
       -I_{1/3}\bigg({2\over3}z^{3/2}\bigg)\Bigg]
  ={1\over\pi}\bigg({z\over3}\bigg)^{1/2}
       K_{1/3}\bigg({2\over3}z^{3/2}\bigg)
  \qquad
  \tag\BB.\num\\    \global\plus
  \Bi(z)&=\bigg({z\over3}\bigg)^{1/2}
  \Bigg[I_{-1/3}\bigg({2\over3}z^{3/2}\bigg)
       +I_{1/3}\bigg({2\over3}z^{3/2}\bigg)\Bigg]\enspace.
  \tag\BB.\num\endalign$$
Then we obtain
\plus
$$\multline
  G(x'',x';E)={\pi\over\hbar}\bigg({4m^2\over\hbar k}\bigg)^{1/3}
  \Ai\left[\bigg(x_>-{E\over k}\bigg)
       \bigg({2mk\over\hbar^2}\bigg)^{1/3}\right]
  \\   \times
  \left\{\Bi\left[\bigg(x_<-{E\over k}\bigg)
       \bigg({2mk\over\hbar^2}\bigg)^{1/3}\right]
  -\sqrt{3}\Ai\left[\bigg(x_<-{E\over k}\bigg)
       \bigg({2mk\over\hbar^2}\bigg)^{1/3}\right]\right\}\enspace.
  \endmultline
  \tag\BB.\num$$

%----------------------------------------------------------------------
%                          END OF FILEB
%----------------------------------------------------------------------

\newpage\noindent
{\bf References}
\baselineskip=12pt
\bigskip
\eightpoint
\eightrm
\def\refno{\item}
%----------------------------------------------------------------------
\refno{[\AGHH]}
S.Albeverio, F.Gesztesy, R.J.H\o egh-Krohn and H.Holden,
Solvable Models in Quantum Mechanics,
Springer Verlag, Berlin, 1988
%-----------------------------------------------------------------------
\refno{[\ANIKZ]}
V.C.Aguilera-Navarro, H.Iwamoto, E.L.Koo and A.H.Zimerman,
Quantum-Mechanical Solution for the Double Oscillator in a Box;
{\it Nuovo Cimento} {\bf B 62} (1981) 91
%-----------------------------------------------------------------------
\refno{[\AGS]}
J.-P.Antoine, F.Gesztesy and J.Shabani,
Exactly Solvable Models of Sphere Interactions in Quantum Mechanics;
{\it J.Phys.A: Math.Gen.}\ {\bf 20} (1987) 3687
%-----------------------------------------------------------------------
\refno{[\BAVE]}
V.L.Bakhrakh and S.I.Vetchinkin,
Green's Functions of the Schr\"odinger Equation for the Simplest
Systems;
{\it Theor.Math.Phys.}\ {\bf 6} (1971) 283
%-----------------------------------------------------------------------
\refno{[\BVK]}
V.L.Bakhrakh, S.I.Vetchinkin and S.V.Khristenko,
Green's Function of a Multidimensional Isotropic Harmonic Oscillator;
{\it Theor.Math.Phys.}\ {\bf 12} (1972) 776
%-----------------------------------------------------------------------
\refno{[\BAU]}
D.Bauch,
The Path Integral for a Particle Moving in a $\delta$-Function
Potential;
{\it Nuovo Cimento} {\bf B 85} (1985) 118
%-----------------------------------------------------------------------
\refno{[\BHR]}
M.Bordag, D.Henning and D.Robaschik,
Quantum Field Theory With External Potentials Concentrated on Planes;
{\it J.Phys.A: Math.Gen.}\ {\bf 25} (1992) 4483
%-----------------------------------------------------------------------
\refno{[\CIWI]}
P.Y.Cai, A.Inomata and R.Wilson,
Path-Integral Treatment of the Morse Oscillator;
{\it Phys.Lett.}\ {\bf A 96} (1983) 117
%-----------------------------------------------------------------------
\refno{[\CAR]}
M.Carreau,
The Functional Integral for a Free Particle on a Half-Plane;
{\it J.Math.Phys.}\ {\bf 33} (1992) 4139
%-----------------------------------------------------------------------
\refno{[\CFG]}
M.Carreau, E.Farhi and S.Gutmann,
Functional Integral for a Free Particle in a Box;
{\it Phys.Rev.}\ {\bf D 42} (1990) 1194
%-----------------------------------------------------------------------
\refno{[\CCH]}
L.Chetouani, A.Chouchaoui and T.F.Hammann,
Path Integral Solution for a Particle Confined in a Region;
{\it J.Math.Phys.}\ {\bf 31} (1990) 838
%-----------------------------------------------------------------------
\refno{[\CHc]}
L.Chetouani and T.F.Hammann,
Coulomb Green's Function, in an $n$-Dimensional Euclidean Space;
{\it J.Math.Phys.}\ {\bf 27} (1986) 2944
%-----------------------------------------------------------------------
\refno{[\CMS]}
T.E.Clark, R.Menikoff and D.H.Sharp,
Quantum Mechanics on the Half-Line Using Path Integrals;
{\it Phys.Rev.}\ {\bf D 22} (1980) 3012
%-----------------------------------------------------------------------
\refno{[\COFRI]}
A.Consortini and B.R.Frieden,
Quantum-Mechanical Solution for the Simple Harmonic Oscillator in a
Box;
{\it Nuovo Cimento} {\bf B 35} (1976) 153
%-----------------------------------------------------------------------
\item{[\DMN]}
V.V.Dodonov, V.I.Man'ko and D.E.Nikonov,
Exact Propagators for Time-Dependent Coulomb, Delta and Other
Potentials;
{\it Phys.Lett.}\ {\bf A 162} (1992) 359
%----------------------------------------------------------------------
\refno{[\DURa]}
I.H.Duru,
Morse-Potential Green's Function With Path Integrals;
{\it Phys.Rev.}\ {\bf D 28} (1983) 2689
%-----------------------------------------------------------------------
\refno{[\DURd]}
I.H.Duru,
On the Path Integral for the Potential $V=ar^{-2}+br^2$;
{\it Phys.Lett.}\ {\bf A 112} (1985) 421
%-----------------------------------------------------------------------
\refno{[\DURe]}
I.H.Duru,
On the Path Integrations for the Wood-Saxon and Related Potentials;
{\it Phys.Lett.}\ {\bf A 119} (1986) 163
%-----------------------------------------------------------------------
\refno{[\DURf]}
I.H.Duru,
Quantum Treatment of a Class of Time-Dependent Potentials;
{\it J.Phys.A: Math.Gen.}\ {\bf 22} (1989) 4827
%-----------------------------------------------------------------------
\refno{[\EMOTa]}
A.Erd\'elyi, W.Magnus, F.Oberhettinger and F.G.Tricomi (eds.),
Higher Transcendental Functions Vol.II,
McGraw Hill, New York, 1955
%-----------------------------------------------------------------------
\refno{[\EMOTb]}
A.Erd\'elyi, W.Magnus, F.Oberhettinger and F.G.Tricomi (eds.),
Tables of Integral Transforms, Vol.I,
McGraw Hill, New York, 1954
%-----------------------------------------------------------------------
\refno{[\FALK]}
L.A.Falkovsky,
Density Attenuation of Surface Magnetic States;
{\it Sov.Phys.JETP} {\bf 31} (1970) 981
%-----------------------------------------------------------------------
\refno{[\FAKL]}
L.A.Falkovsky and S.Klama,
Broadening of Electron States in a Thin Film Within a Rough Surface
in an External Longitudinal Magnetic Field;
{\it J.Phys.C: Solid State Phys.}\ {\bf 20} (1987) 1751
%-----------------------------------------------------------------------
\refno{[\FARI]}
J.R.Farias,
Simple Exact Solutions of the One-Dimensional Schr\"odinger Equation
with Continuous Plus $\delta$-Function Potentials of Arbitrary
Position and Strength;
{\it Phys.Rev.}\ {\bf A 22} (1980) 765
%-----------------------------------------------------------------------
\refno{[\FH]}
R.P.Feynman and A.Hibbs, Quantum Mechanics and Path Integrals,
McGraw Hill, New York, 1965
%----------------------------------------------------------------------
\refno{[\FLU]}
S.Fl\"ugge,
Practical Quantum Mechanics, Vol.I;
Die Grundlagen der mathematischen Wis\-sen\-schaf\-ten Vol.177
Springer-Verlag, Berlin, 1977
%-----------------------------------------------------------------------
\refno{[\GASCH]}
B.Gaveau and L.S.Schulman,
Explicit Time-Dependent Schr\"odinger Propagators;
{\it J. Phys.A: Math.Gen.}\ {\bf 19} (1986) 1833
%-----------------------------------------------------------------------
\refno{[\GOO]}
M.J.Goovaerts,
Path-Integral Evaluation of a Nonstationary Calogero Model;
{\it J.Math.Phys.}\ {\bf 16} (1975) 720
%-----------------------------------------------------------------------
\refno{[\GOBR]}
M.J.Goovaerts and F.Broeckx,
Analytic Treatment of a Periodic $\delta$-Function Potential in the
Path Integral Formalism;
{\it SIAM J.Appl.Math.}\ {\bf 45} (1985) 479
%-----------------------------------------------------------------------
\refno{[\GRA]}
I.S.Gradshteyn and I.M.Ryzhik,
Table of Integrals, Series, and Products,
Academic Press, New York, 1980
%-----------------------------------------------------------------------
\refno{[\GROb]}
C.Grosche,
The Path Integral on the Poincar\'e Upper Half-Plane With
a Magnetic Field and for the Morse Potential;
{\it Ann.Phys.(N.Y.)} {\bf 187} (1988) 110
%-----------------------------------------------------------------------
\refno{[\GROc]}
C.Grosche,
The Path Integral on the Poincar\'e Disc, the Poincar\'e Upper
Half-Plane and on the Hyperbolic Strip;
{\it Fortschr.Phys.}\ {\bf 38} (1990) 531
%----------------------------------------------------------------------
\refno{[\GROe]}
C.Grosche,
Path Integral Solution of a Class of Potentials Related to the
P\"oschl-Teller Potential;
{\it J.Phys.A: Math.Gen.}\ {\bf 22} (1989) 5073
%-----------------------------------------------------------------------
\refno{[\GROh]}
C.Grosche,
Path Integrals for Potential Problems With $\delta$-Function
Perturbation;
{\it J.Phys.A: Math.Gen.}\ {\bf 23} (1990) 5205
%-----------------------------------------------------------------------
\refno{[\GROm]}
C.Grosche,
Coulomb Potentials by Path-Integration;
{\it Fortschr.Phys.}\ {\bf 40} (1992) 695
%-----------------------------------------------------------------------
\refno{[\GROu]}
C.Grosche,
Selberg Trace-Formul\ae\ in Mathematical Physics;
{\it Trieste pre\-print}, SISSA/\-177/\-92/\-FM,
to appear in: {\it Proceedings of the Workshop ``From Classical to
Quantum Chaos (1892-1992)'', 21-24 July, 1992, Trieste}, World Scientific,
Singapore; eds.: G.\ Dell'Antonio, S.\ Fantoni and V.\ R.\ Manfredi
%-----------------------------------------------------------------------
\refno{[\GROw]}
C.Grosche,
Path Integral Solution of a Class of Explicitly Time-Dependent
Potentials;
{\it Trieste preprint}, SISSA/2/93/FM
%-----------------------------------------------------------------------
\refno{[\GRSa]}
C.Grosche and F.Steiner,
The Path Integral on the Poincar\'e Upper Half
Plane and for Liouville Quantum Mechanics;
{\it Phys.Lett.}\ {\bf A 123} (1987) 319
%-----------------------------------------------------------------------
\refno{[\GRSb]}
C.Grosche and F.Steiner,
Path Integrals on Curved Manifolds;
{\it Zeitschr.Phys.}\ {\bf C 36} (1987) 699
%-----------------------------------------------------------------------
\refno{[\GRSf]}
C.Grosche and F.Steiner,
Classification of Solvable Feynman Path Integrals;
{\it DESY preprint} DESY 92-189, to appear in the {\it Proceedings of
the ``Fourth International Conference on Path Integrals from $meV$ to
$MeV$'', May 1992, Tutzing, Germany}, World Scientific, Singapore
%----------------------------------------------------------------------
\refno{[\GRSg]}
C.Grosche and F.Steiner,
Table of Feynman Path Integrals;
to appear in: {\it Springer Tracts in Modern Physics}
%-----------------------------------------------------------------------
\refno{[\INO]}
A.Inomata,
Remarks on the Experiment of Winding Number Dependence of the
Aharonov-Bohm Effect;
{\it Phys.Lett.}\ {\bf A 95} (1983) 176
%-----------------------------------------------------------------------
\refno{[\INOSa]}
A.Inomata and V.A.Singh,
Path Integrals With a Periodic Constraint: Entangled Strings;
{\it J.Math.Phys.}\ {\bf 19} (1978) 2318
%-----------------------------------------------------------------------
\refno{[\INOSb]}
A.Inomata and V.A.Singh,
Path Integrals and Constraints: Particle in a Box;
{\it Phys.Lett.}\ {\bf A 80} (1980) 105
%-----------------------------------------------------------------------
\refno{[\JAKL]}
W.Janke and H.Kleinert,
Summing Paths for a Particle in a Box;
{\it Lett.Nuovo Cimento} {\bf 25} (1979) 297
%-----------------------------------------------------------------------
\refno{[\KLRO]}
S.Klama and U.R\"ossler,
The Green's Function of Confined Electrons in an External Magnetic
Field;
{\it Ann.Physik} {\bf 1} (1992) 460
%-----------------------------------------------------------------------
\refno{[\KLE]}
H.Kleinert,
Path Integrals in Quantum Mechanics, Statistics and Polymer Phys\-ics,
World Scientific, Singapore, 1990
%-----------------------------------------------------------------------
\refno{[\KLEMUS]}
H.Kleinert and I.Mustapic,
Summing the Spectral Representations of P\"oschl-Teller and
Rosen-Morse Fixed-Energy Amplitudes;
{\it J.Math.Phys.}\ {\bf 33} (1992) 643
%-----------------------------------------------------------------------
\refno{[\KOM]}
I.V.Komarov,
Application of the Short Range Potential in the Calculations of the
Ion-Ion Recombination; {\it Sixth International Conference on the
Physics of Electronics and Atomic Collisions}, Massachusetts, 1969,
abstract of papers, p.1015
%-----------------------------------------------------------------------
\refno{[\LABH]}
S.V.Lawande and K.V.Bhagwat,
Feynman Propagator for the $\delta$-Function Potential;
{\it Phys.Lett.}\ {\bf A 131} (1988) 8
%-----------------------------------------------------------------------
\refno{[\MCG]}
J.B.McGuire,
Study of Exactly Soluble One-Dimensional $N$-Body Problems;
{\it J.Math.Phys.}\ {\bf 5} (1964) 622
%-----------------------------------------------------------------------
\refno{[\MCGHU]}
J.B.McGuire and C.A.Hurst,
The Scattering of Three Impenetrable Particles in One Dimension;
{\it J.Math.Phys.}\ {\bf 13} (1972) 1595
%-----------------------------------------------------------------------
\refno{[\MOS]}
W.Magnus, F.Oberhettinger and R.P.Soni,
Formulas and Theorems for the Special Functions of Mathematical Physics,
Springer-Verlag, Berlin, 1966
%-----------------------------------------------------------------------
\refno{[\MACR]}
J.L.Marin and S.A.Cruz,
On the Harmonic Oscillator Inside an Infinite Potential Well;
{\it Amer.J. Phys.}\ {\bf 56} (1988) 1134
%-----------------------------------------------------------------------
\refno{[\PAKSa]}
K.Pak and I.S\"okmen,
A New Exact Path Integral Treatment of the Coulomb and the Morse
Potential Problems;
{\it Phys.Lett.}\ {\bf A 100} (1984) 327
%-----------------------------------------------------------------------
\refno{[\PI]}
D.Peak and A.Inomata,
Summation Over Feynman Histories in Polar Coordinates;
{\it J.Math.Phys.}\ {\bf 10} (1969) 1422
%-----------------------------------------------------------------------
\refno{[\SCHU]}
L.S.Schulman,
Techniques and Applications of Path Integration,
John Wiley \&\ Sons, New York, 1981
%-----------------------------------------------------------------------
\refno{[\SIEB]}
M.Sieber,
The Hyperbola Billiard: A Model for the Semiclassical Quantization of
Chaotic Systems;
{\it DESY preprint}, DESY 91-030
%-----------------------------------------------------------------------
\refno{[\STEc]}
F.Steiner,
Exact Path Integral Treatment of the Hydrogen Atom;
{\it Phys.Lett.}\ {\bf A 106} (1984) 363
%-----------------------------------------------------------------------
\refno{[\STEa]}
F.Steiner,
Path Integrals in Polar Co-ordinates From $eV$ to $GeV$;
in ``Bielefeld Encounters in Physics and Mathematics VII; Path
Integrals From $meV$ to $MeV$'', 1985, eds.: M.\ C.\ Gutzwiller et al.\
World Scientific, Singapore, 1986, p.335
%-----------------------------------------------------------------------
\refno{[\STEb]}
F.Steiner,
On Selberg's Zeta Function for Compact Riemann Surfaces;
{\it Phys.Lett.} {\bf 188B} (1987), 447;
%-----------------------------------------------------------------------
Quantum Chaos and Geometry;
in {\it Recent Developments in Mathematical Physics}, Conference
Schladming 1987, eds.: H.\ Mitter, L.\ Pittner,
Springer-Verlag, Berlin, 1987, p.305
%-----------------------------------------------------------------------
\refno{[\VAW]}
R.Vawter,
Effects of Finite Boundaries on a One-Dimensional Harmonic Oscillator;
{\it Phys.Rev.}\ {\bf 174} (1968) 749
%-----------------------------------------------------------------------
\refno{[\WIE]}
F.W.Wiegel,
Path Integrals With Topological Constraints: Aharonov-Bohm Effect and
Polymer Entanglements;
{\it Physica} {\bf A 109} (1981) 609
%-----------------------------------------------------------------------
\refno{[\ZHCH]}
S.K.Zhdanov and A.S.Chikhachev,
Particle in a Field of Dispersing $\delta$-Potentials;
{\it Sov.Phys.Dokl.}\ {\bf 19} (1975) 696
%-----------------------------------------------------------------------
%                          END OF FILER
%----------------------------------------------------------------------

%----------------------------------------------------------------------
%                        TOTAL END OF FILE
%----------------------------------------------------------------------

\enddocument